# Optimized Designs for High-Efficiency Particle Sorting in Serpentine Microfluidic Channels

**Sayan Karmakar[1,2], Anish Pal[2], Sourav Sarkar[3*], Achintya Mukhopadhyay[3]**

[1]Department of Civil Engineering, Jadavpur University, Kolkata, India
[2]Department of Mechanical and Industrial Engineering, University of Illinois at Chicago, Chicago, IL, USA
[3]Departement of Mechanical Engineering, Jadavpur University, Kolkata, India
*Corresponding author email: souravsarkar.mech@jadavpuruniversity.in

**ABSTRACT:**

Efficient particle sorting in microfluidic systems is vital for advancements in biomedical diagnostics and industrial applications. This study numerically investigates particle migration and passive sorting in symmetric serpentine microchannels, leveraging inertial and centrifugal forces for label-free, high-throughput separation. Using a two-dimensional numerical model, particle dynamics were analyzed across varying flow rates, diameter ratios (1.2, 1.5, and 2), and channel configurations. The optimized serpentine geometry achieved particle separation efficiencies exceeding 95% and throughput greater than 99%.A novel scaling framework was developed to predict the minimum number of channel loops required for efficient sorting. Additionally, the robustness of the proposed scaling framework is demonstrated by its consistency with findings from previous studies, which exhibit the same trend as predicted by the scaling laws, underscoring the universality and reliability of the model. Additionally, the study revealed the significant influence of density ratio (α) on sorting efficiency, where higher α values enhanced separation through amplified hydrodynamic forces. Optimal flow rates tailored to particle sizes were identified, enabling the formation of focused particle streaks for precise sorting. However, efficiency declined beyond these thresholds due to particle entrapment in micro-vortices or boundary layers. This work provides valuable insights and design principles for developing compact, cost-effective microfluidic systems, with broad applications in biomedical fields like cell sorting and pathogen detection, as well as industrial processes requiring precise particle handling

## NOMENCLATURE:

| Symbol | Description | Unit |
|---|---|---|
| $a, d_p$ | Particle diameter | m |
| $\tilde{a}$ | Dimensionless particle diameter | |
| $a_{base}$ | Base particle diameter | m |
| $AR$ | Aspect ratio | -- |
| $D_h$ | Hydraulic diameter | m |
| $f_c$ | Lift coefficient | -- |
| $F_A$ | Archimedes force | N |
| $F_b$ | External body force in Eulerian phase | $N\ m^{-3}$ |
| $F_{Ba}$ | Basset force | N |
| $F_{Cent}$ | Centrifugal force on particle in Lagrangian phase | N |
| $F_D$ | Drag force on particle in Lagrangian phase | N |
| $F_L$ | Lift force on particle in Lagrangian phase | N |
| $F_{VM}$ | Virtual mass force | N |
| $\tilde{F}$ | Nondimensional force | -- |
| $k_{scale}$ | Scaling parameter | -- |
| $l$ | Length of a serpentine loop | m |
| $L_e$ | Entrance length for fluid flow within a microchannel | m |
| $L_{min}$ | Minimum length required for particle focusing | m |
| $L_r$ | Distance travelled by a particle in radial direction of the channel | m |
| $L_t$ | Distance travelled by a particle in tangential direction of the channel | m |
| $m_p$ | Mass of a particle | kg |
| $\tilde{m}_p$ | Nondimensional mass of particle | |
| $n$ | Particle diameter ratio | -- |
| $N_{loop}$ | Loop number | -- |
| $N_{min}$ | Minimum number of loops required for particle focusing | -- |
| $N_{simulation}$ | Minimum number of loops for sorting from simulation | -- |
| $p$ | Static pressure of the Eulerian phase | $N\,m^{-2}$ |
| $\tilde{p}$ | Non-dimensional pressure | |
| $Q$ | Channel flow rate | $m^3\,s^{-1}$ |
| $r$ | Radius of curvature | m |
| $r_p$ | Particle radius | m |
| $Re_c$ | Channel Reynolds number | -- |
| $Re_p$ | Particle Reynolds number | -- |
| $Re_r$ | Relative particle Reynolds number | -- |
| $Stk$ | Stokes number | -- |
| $t$ | Time | |
| $\tilde{t}$ | Non-dimensional time | |
| $u$ | Fluid velocity in Eulerian phase | $m\,s^{-1}$ |
| $\tilde{u}$ | Dimensionless velocity | |
| $u_p$ | Velocity of a particle in Lagrangian phase | $m\,s^{-1}$ |
| $U_f$ | Average fluid velocity | $m\,s^{-1}$ |
| $U_L$ | Lateral migration velocity of particle | $m\,s^{-1}$ |
| $U_m$ | Maximum fluid velocity at inlet | $m\,s^{-1}$ |
| $v_{fr}$ | Radial fluid velocity | $m\,s^{-1}$ |
| $v_{pr}$ | Radial particle velocity | $m\,s^{-1}$ |
| $v_{pt}$ | Tangential particle velocity | $m\,s^{-1}$ |
| $v_{prr}$ | Relative radial particle velocity | $m\,s^{-1}$ |
| $w$ | Width of the microchannel | m |
| $\alpha$ | Particle density ratio | |
| $\delta$ | Particle focusing efficiency | -- |
| $\varepsilon$ | Ratio of Basset force to drag force | -- |
| $\eta$ | Particle separation efficiency | -- |
| $\eta_{Cent}$ | Effect of Centrifugal force on focusing | -- |
| $\mu$ | Dynamic viscosity of the Eulerian phase | $kg\,m^{-1}\,s^{-1}$ |
| $\nu$ | Kinematic viscosity of the Eulerian phase | $m^2\,s^{-1}$ |
| $\xi$ | Confinement ratio for a microchannel | -- |
| $\rho_f$ | Density of the Eulerian phase | $kg\,m^{-3}$ |
| $\rho_p$ | Particle density | $kg\,m^{-3}$ |
| $\tau_p$ | Particle response time | s |
| $\tilde{\tau}_p$ | Nondimensional particle response time | |
| $\sim$ | Mark for non-dimensional parameters | -- |

# 1. INTRODUCTION:

Cells, as the fundamental units of life, carry essential information crucial for detecting and predicting various diseases. The analysis of both healthy and damaged cells is pivotal for clinical applications and biological research, as these cells often coexist in bodily fluids. Therefore, the development of efficient, scalable separation techniques for isolating these cells has become crucial[1–5].

Particle separation methods are generally classified into active and passive sorting techniques. Active sorting methods, such as electrophoresis[6,7], dielectrophoresis (DEP)[8], magnetophoresis (MP)[9], and acoustophoresis (AP)[10], rely on the interaction between external force fields and hydrodynamic forces. These techniques, while effective, often require bulky, expensive instrumentation, are prone to contamination, and can potentially damage particles during sorting. On the other hand, passive sorting techniques like mechanical filtering[11], pinched flow fractionation (PFF)[12], deterministic lateral displacement (DLD)[13], and inertial microfluidics[14] depend solely on the natural behavior of fluid flow and channel geometry. Passive techniques, particularly inertial microfluidics, are gaining attention due to their potential for label-free, high-precision, high-throughput particle sorting without the need for damaging immunolabelling procedures[15–18]. Inertial microfluidics offers several advantages: reduced system size, lower costs, faster sample processing, and minimal human interaction, making it an attractive alternative for cell sorting in medical diagnostics and biological research[19–25].

Inertial microfluidics has revolutionized particle sorting by utilizing the complex interplay of hydrodynamic forces to manipulate particles based on their physical properties[26–30]. This technique leverages the natural flow characteristics within microchannels to achieve efficient separation based on size, deformability, and other intrinsic properties[31–39]. These systems allow for high-throughput sorting with enhanced precision, significantly improving the ability to isolate target cells from heterogeneous populations[40–42]. Furthermore, inertial microfluidics eliminates the need for complex labelling and offers significant improvements in sorting speed and efficiency compared to traditional methods[43,44].

In this study, we focus on particle migration within serpentine microchannels, a widely studied geometry in inertial microfluidics. Previous studies, such as those by Zhang et al.[45,46], have explored particle migration and focusing within symmetric serpentine microchannels. However, these works neglected the impact of inertial lift forces on particle behavior, assuming blockage ratios $(a/D_h)$ less than $0.07^{14}$. In contrast, the present study considers particle combinations with blockage ratios exceeding 0.07, where inertial lift forces play a critical role in focusing and sorting efficiency. We demonstrate that neglecting inertial lift forces leads to poor sorting efficiency, while incorporating these forces significantly enhances particle focusing within the microchannel.

In the present study, a 2D simulation has been used to investigate particle migration and focusing in a serpentine microchannel geometry of width 200 μm. Here, the fluid flow has been simulated in the Eulerian phase, while utilizing Discrete Phase Modeling (DPM), the motion of each particle is tracked in a Lagrangian framework. Consequently, sorting efficiency and throughput have been analyzed under varying flow rates to understand their effects on the performance of the sorting process.

This work investigates several key parameters, including particle diameter, diameter ratio (n), channel s the competing forces acting on the particles within the microchannel.

The study also addresses the impact of the number of loops in the serpentine microchannel on sorting efficiency, providing a scaling analysis to determine the minimum number of loops required for effective separation. These findings were validated through numerical simulations, offering a deeper understanding of the role of inertial forces and the design of microfluidic channels in achieving high-performance particle sorting.

By leveraging the power of inertial microfluidics, this study presents a significant advancement in label-free particle sorting. The results highlight the potential for improving sorting efficiency and throughput, providing a promising tool for medical diagnostics, biological research, and pathogen detection. The ability to manipulate particles based on their intrinsic properties without the need for labeling or complex equipment is a major step forward in the field of microfluidics and nanofluidics.

## 2. **PROBLEM FORMULATION:**

The main focus of the present study is to deal with the inertial sorting of different-sized particles, through a serpentine microchannel, of AR = 0.25. As can be visualized in Fig. 1, the microchannel has a width (w) of 200 $\mu m$ and a depth (H) of 50 $\mu m$. For the present study, varying particle sizes, for n = 1.2, 1.5, 2 has been considered, where 7.5 $\mu m$ is the base particle diameter. Under different flow rates (Q) their sorting efficiency and throughput has been determined, at the bifurcated outlet. Particles entering into the channel from the left pass through a number of serpentine loops, and ultimately get separated at the bifurcated channel outlet.

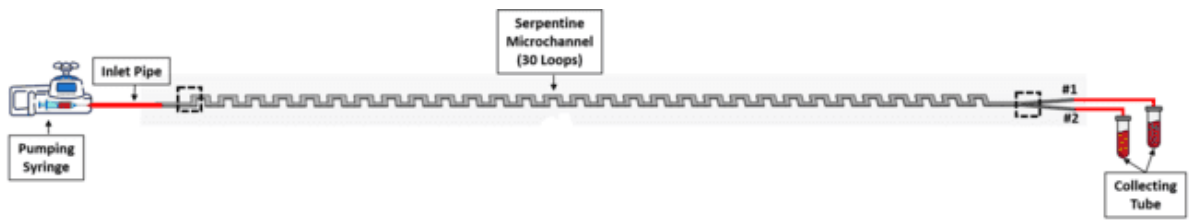

**Figure 1:** Symmetric serpentine microchannel geometry

## 3. **MODELLING:**

### 3.1. **Model Geometry:**

This paper presents an inertial focusing in a two-dimensional microfluidic channel, initially having 30 loops and a width of 200 $\mu m$, as shown in Fig. 2. The outlet is bifurcated for a separate collection of particles. For modeling purpose, a 2D geometry has been considered here. Comparisons between 2D and 3D flow profiles have also been performed. To increase the readability of the paper, that has been included in appendix, A.1. At the inlet, the velocity varies in accordance with the channel Reynolds number and flow rate. Both the inlet and outlet are open to normal atmospheric pressure. (The pumping syringe is shown only for visualisation purpose)

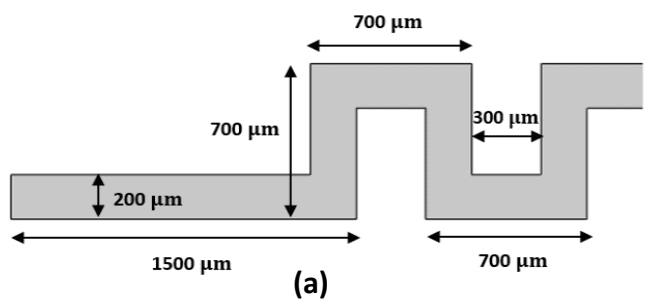

(a)

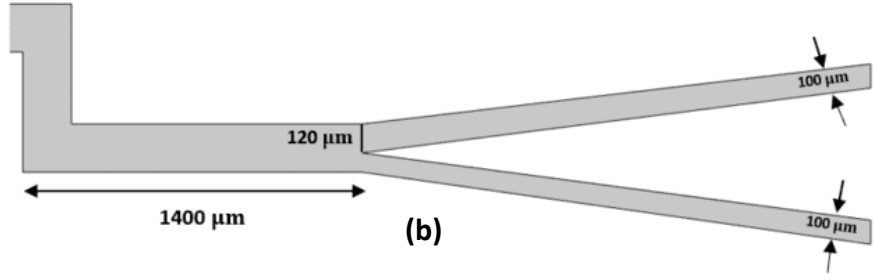

**Figure 2:** (a) Inlet region dimensions (b) Bifurcated outlet channel dimensions

### 3.2. Mathematical Modelling:

The Navier-Stokes Equations for the conservation of mass and momentum are solved for steady and incompressible flow. Often blood samples and other bodily fluids, shows non-Newtonian behavior. In experimental microfluidics for the separation of Circulating Tumor Cells (CTCs), Red Blood Cell (RBC), and White Blood Cell (WBC), the cell samples are mixed with external solvents and diluted[47,48]. So, the intrinsic properties of the fluid like viscosity, and density are altered. Zhou et al.[48] in their works also diluted the blood sample using 10% Fetal Bovine Serum (FBS). Hence from a practical point of view for simulation and calculation, the background fluid within the microchannel is considered to be Newtonian[49,50]. Due to the small dimensions of a microchannel, relatively low velocity, and consequently low Reynolds number ($Re_c$), the flow field was solved using a laminar flow module by an Eulerian approach. The particles are simulated within the fluid phase using the Discrete phase modelling (DPM), and each particle was tracked using the Lagrangian framework. For low relative Reynolds numbers, Stokes's drag law which allows for wall correction was used to solve the drag force. This wall correction in the drag force includes the near-wall effects on the particles.

**Governing equations for the Eulerian (fluid) phase**,

Continuity equation,

$$\frac{\partial u_j}{\partial x_j} = 0 \tag{1}$$

Navier Stokes momentum equations,

$$\frac{\partial}{\partial t}(\rho_f u_i) + \frac{\partial}{\partial x_j}(\rho_f u_i u_j) = -\frac{\partial p}{\partial x_i} + \mu \frac{\partial^2 u_i}{\partial x_j^2} \tag{2}$$

**Governing equations for the Lagrangian phase**,

Equation of motion of particles,

$$\frac{d}{dt}(m_p u_p) = F_D + F_L \tag{3}$$

Relative particle Reynolds number in fluid,

$$Re_r = \frac{\rho_f a |u_{fr} - u_{pr}|}{\mu} \tag{4}$$

Particle Reynolds number,

$$Re_p = Re_c \frac{a^2}{D_h^2} = \frac{U_m a^2}{\nu D_h} \tag{5}$$

Particle velocity response time for spherical particles in laminar flow,

$$\tau_p = \frac{\rho_p a^2}{18\mu} \tag{6}$$

Stokes number,

$$Stk = \frac{\tau_p U_m}{a} \tag{7}$$

Drag force from Stokes drag law,

$$F_D = \left(\frac{1}{\tau_p}\right) m_p (u_{\text{fr}} - u_{\text{pr}}) = 3\pi\mu a (u_{\text{fr}} - u_{\text{pr}}) \tag{8}$$

Saffman Lift Force,

$$F_L = 1.615 a^2 L_v \sqrt{\mu \rho_f \frac{|u - u_p|}{|L_v|}} \tag{9}$$

$$L_v = (u - u_p) \times [\nabla \times (u - u_p)] \tag{10}$$

Now, for nondimensionalising the equations, corresponding non-dimensional parameters are defined below,

$$\tilde{t} = \frac{t}{t_c} = \frac{t U_m}{D_h}, \quad \tilde{x} = \frac{x}{D_h}, \quad \tilde{a} = \frac{a}{D_h}, \quad \tilde{u} = \frac{u}{U_m}, \quad \tilde{p} = \frac{p}{\rho_f U_m^2}, \quad \tilde{m}_p = \frac{m_p}{\rho_f D_h^3}, \quad \tilde{F} = \frac{F}{\rho_f U_m^2 D_h^2}$$

$$\alpha = \frac{\rho_p}{\rho_f}, \quad \tilde{\tau}_p = \frac{\tau_p}{t_c} = \frac{\rho_p a^2}{18\mu} \cdot \frac{U_m}{D_h} = \frac{\alpha \tilde{a}^2 Re_c}{18}, \quad Re_c = \frac{\rho_f U_m D_h}{\mu}, \quad Stk = \frac{\alpha \tilde{a} Re_c}{18}$$

**Governing equations for the Eulerian (fluid) phase**,

Continuity equation,

$$\frac{\partial \tilde{u}_j}{\partial \tilde{x}_j} = 0 \tag{11}$$

Navier Stokes momentum equations,

$$\frac{\partial}{\partial \tilde{t}}(\tilde{u}_i) + \frac{\partial}{\partial \tilde{x}_j}(\tilde{u}_i \tilde{u}_j) = -\frac{\partial \tilde{p}}{\partial \tilde{x}_i} + \frac{1}{Re_c}\frac{\partial^2 \tilde{u}_i}{\partial \tilde{x}_j^2} \tag{12}$$

**Governing equations for the Lagrangian phase**,

Equation of motion of particles,

$$\frac{d}{d\tilde{t}}(\tilde{m}_p \tilde{u}_p) = \tilde{F}_D + \tilde{F}_L \tag{13}$$

Drag force from Stokes drag law,

$$\tilde{F}_D = \frac{3\pi \tilde{a}}{Re_c}(\tilde{u}_{fr} - \tilde{u}_{pr}) = \frac{\pi}{6}\frac{\tilde{a}^2 \alpha}{Stk}(\tilde{u}_{fr} - \tilde{u}_{pr}) \tag{14}$$

Saffman Lift Force,

$$\tilde{F}_L = 6.46 \tilde{a}^2 \tilde{L}_v \frac{\sqrt{|\tilde{u} - \tilde{u}_p|}}{\sqrt{Re_c |\tilde{L}_v|}} \tag{15}$$

$$|\tilde{L}_v| = \frac{D_h}{U_m^2}|L_v| \tag{16}$$

Here $\rho_f$ is the density of the fluid, $p$ is the pressure, $u$ is the velocity and $\mu$ is the dynamic viscosity. The fluid, mimicking diluted blood sample is considered with an average density $\rho_f = 994\ kg/m^3$ and dynamic viscosity $\mu = 0.0004\ Pa \cdot s$ [51,52]. Different cellular particles are simulated here, with an average density $\rho_p = 1110\ kg/m^3$ [53].

To track the particle locations, Eq. 4, Newton's second law of motion is employed. Here $m_p$ is the mass of the particle, $v_p$ is the particle velocity. $F_D$ is the integrated drag force and $F_L$ is the inertial Saffman lift force [54]. Several forces can drive the particle motion such as drag, gravity, electric, magnetic, lift, and acoustophoretic forces. In this case drag, lift and centrifugal force are acting on the particle.

Apart from these, other forces like virtual mass force, Archimedes force and Basset force may also play a critical role on particle motion within a fluid through a microchannel[55–59]. Since, in the present work, our main focus is efficient and optimized particle sorting, these forces have been neglected owing to their very minor contribution on particle trajectories and sorting efficiency. Appendix, A.5. encompasses the further discussion of the role of these forces on particle sorting.

Here the time-dependent solution is performed using a Eulerian method. For small $Re_p$ and $\frac{a}{D_h} \ll 1$, the particles do not disturb the underlying flow, hence only the hydrodynamic forces alter the particle behavior (see appendix: A.4). Hence a one-way fluid-particle coupling is performed. Thus, the drag force acting on the particles, is calculated from the velocity solution of the laminar flow profile.

### 3.3. Initial and boundary conditions:

A transient time-dependent solution is computed for this numerical study. Initially the flow field is solved, and then the particles are released within the fluid, at a time when the fluid has attained a steady state. Hence, all the particles are released into the flow at $t = 1\ sec$, so that the steady flow regime is achieved before the particle release, as can be seen in Fig. 3(e). Along with this, the particle diameter ratio defined as $n = \frac{d_{large}}{d_{small}}$, (ratio of the diameter of the largest and smallest particle), is also crucial in reflecting the particle behavior which in turn affects the flow in the confined microchannel. For every simulation, only 2 types of particles are present at a time with 1000 particles each. Three different values of n are considered viz. 1.2 (7.5 μm & 9 μm), 1.5 (7.5 μm & 11.25 μm), 2 (7.5 μm & 15 μm). The average size of cellular particles is represented by 7.5 μm, 11.25 μm, and 15 μm[60,61]. Starting from an initial flow rate of 200 μL/min upto to around 600 μL/min, corresponding to each diameter ratio combination, an appropriate flow rate was determined and analyzed. At the microchannel walls, no-slip boundary conditions were applied, and then the calculated flow field was used to trace the particles. Furthermore, for the particle-wall interaction, specular reflection model is incorporated to simulate the particle bounce boundary condition (refer appendix A.6.). At the outlets, zero pressure boundary conditions were imposed.

### 3.4. Numerical Methods:

A Finite Element based commercial CFD Solver, COMSOL Multiphysics, v5.6[62], is used, for this numerical study. Given that the motion in the laminar domain can be identified as a sparse flow of small particles' mass, volume, and density, the drag force feature is required. Due to the small dimensions of a microchannel, relatively low velocity, and consequently low relative Reynolds number ($Re_c$), the flow field was solved using a laminar flow module in COMSOL[62] by an Eulerian approach. The particles were simulated within the fluid phase using the Discrete phase modeling (DPM), and each particle was tracked using the Lagrangian framework.

The Navier-Stokes equations are solved using the Finite Element Method (FEM) and the P1+P1 scheme is selected to discretize the velocity and pressure fields. In COMSOL, the default discretization scheme for laminar flow is P1+P1, where 1 stands for linear first-order elements for both velocity and pressure components. Similarly, COMSOL Multiphysics, also avails the choice for higher order velocity and pressure elements, through other discretization schemes such as P2+P2 and P3+P3. Unlike higher-order elements, linear elements are computationally cheaper and are also less prone to introduce spurious oscillations, thereby improving the numerical robustness. In the present numerical study, the chosen relative tolerance criteria for convergence is $10^{-5}$.

### 3.5. Meshing and Grid Independence Study:

A 2D unstructured triangular-dominant mesh is used for the discretization of the computational domain. Figure 3(a) depicts the overall meshing of the whole serpentine microchannel geometry. To perform the grid-independent study, a microchannel of width 200 μm, having 30 loops is chosen. Now 2 particular probe point locations are chosen within the 15$^{th}$ loop as shown in Fig. 3(b) for studying the fluid velocity. Three simulations were done under 3 different mesh densities to analyze the fluid velocity component ($u$), at the probe point locations. The grid independence is performed under a flow rate of Q = 260 μL/min, corresponding to $Re_c$ = 86.147. Thus, considering the mesh refinement the probe velocity analysis for three different mesh densities viz. mesh 1 (course - 72970 elements), mesh 2 (medium - 109394 elements), and mesh 3 (fine - 125149 elements) the grid-dependent analysis is done, as shown in Figs. 3(c) and 3(d). As seen from the grid independence comparison, in Table I, there is negligible variation between mesh 2 and 3. Hence computationally cheaper mesh 2 is chosen for subsequent numerical study.

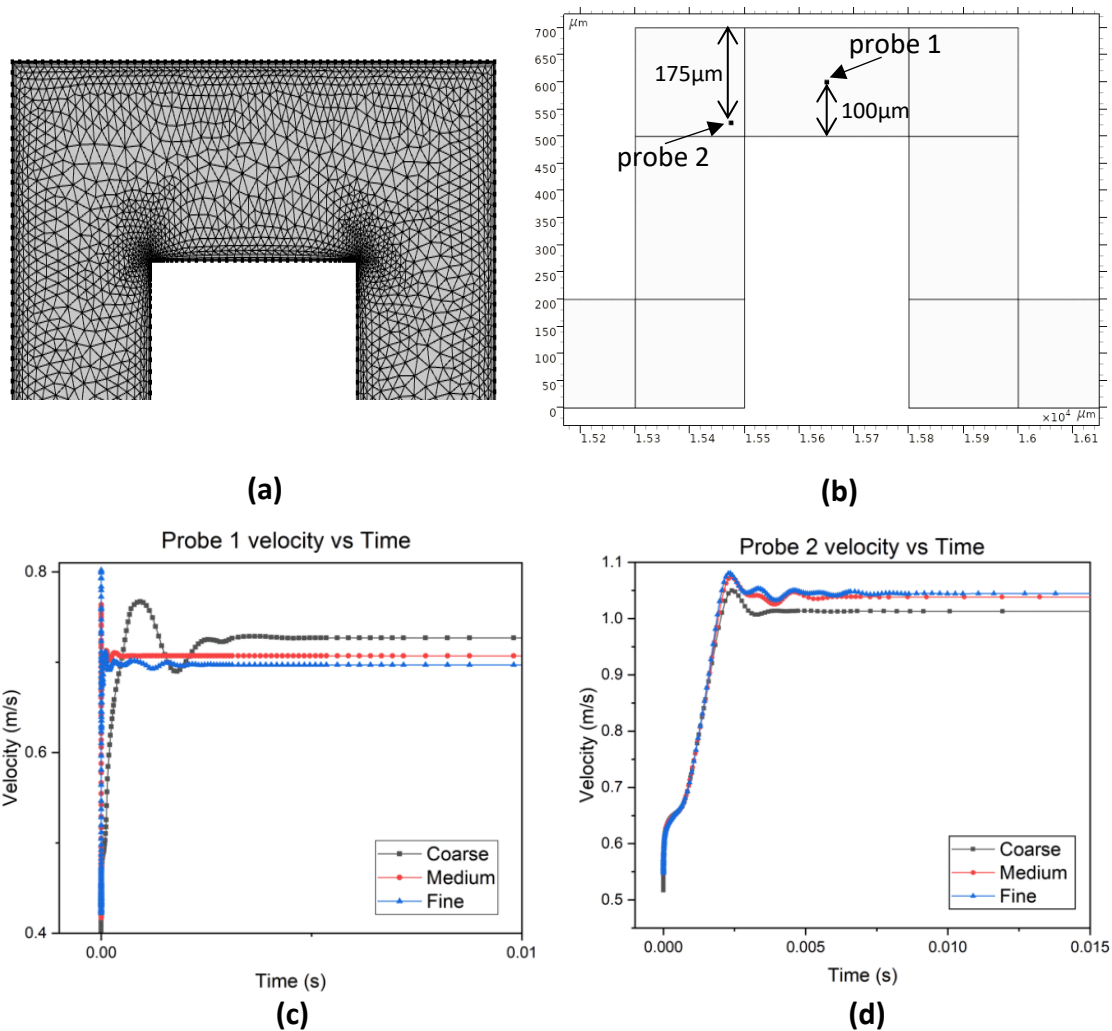

**Figure 3:** (a) Adopted meshing in a particular loop with wall-edge and corner refinement. (b) Point probe locations within the microchannel. (c), (d) Probe 1 & 2 velocity plot for Mesh 1, Mesh 2, and Mesh 3.

**TABLE I:** Grid-independence data:

| Mesh | Number of Elements | Probe 1 | | Probe 2 | |
|---|---|---|---|---|---|
| | | Velocity magnitude (m/s) | Change (%) | Velocity magnitude (m/s) | Change (%) |
| Mesh 1 – Coarse | 72970 | 0.72694 | | 1.01299 | |
| Mesh 2 – Medium | 109394 | 0.70704 | 2.74 | 1.03838 | 2.51 |
| Mesh - Fine | 125149 | 0.69717 | 1.39 | 1.04441 | 0.58 |

### 3.6. Time-Independence study:

Table 2, below depicts the time-independence study in detail. Initially three different time step sizes of $10^{-3}$, $10^{-4}$ and $10^{-5}$ are selected and the corresponding magnitude of the point probe 1 velocity is shown for these time steps.

**TABLE 2:** Showing time-independence data:

| Time step size | The magnitude of probe velocity ($m/s$) | Percentage change from previous time step value (%) |
|---|---|---|
| $10^{-3}$ | 0.706852 | |
| $10^{-4}$ | 0.707043 | 0.027 |
| $10^{-5}$ | 0.707042 | 0.00014 |

The probe velocity and Courant Number (Co) variation plots from Figs. 4(a) and 4(b) against varying time step sizes show, that there is not much considerable variation between the results of either time step size. But the percentage error between the time step $10^{-4}$ and $10^{-5}$ is considerably less than that of $10^{-3}$ and $10^{-4}$. So, for these time-step sizes, the results are practically time-step independent. Hence, the time-step size of $10^{-4}$ is selected for this study as it incurs less computation time.

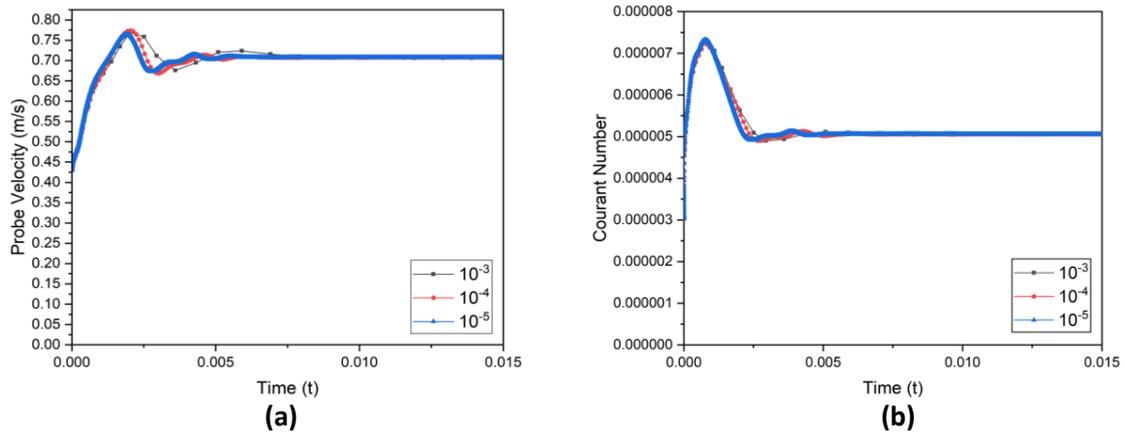

**Figure 4:** (a) Probe point velocity variation plot against different time step sizes. (b) Courant Number (Co) variation plot inside the microchannel domain against different time step sizes.

### 3.7. Model Validation:

Before proceeding with the numerical model for actual study, the laminar flow model has been validated against the experimental data of Ahmad et al.[63] for analyzing microchannel entrance length characteristics, corresponding to different Re values. A square-sectioned, straight microchannel with a hydraulic diameter ($D_h$) of 200 $\mu m$ is employed for determining the entrance length ($L_e$) of the fluid flow, after which the fully developed flow profile is achieved. Here the working fluid was distilled water. The experiments were conducted for varying $Re_c$ values ($Re_c$ < 200) and corresponding to each $Re_c$, a non-dimensional entrance length ($L_e/D_h$) was plotted, as shown in Fig. 5(a). These experimental results from Ahmad et al.[63] are compared with numerical results of the present model, and the match was found to be reasonably good. Hence the laminar flow model is validated.

Subsequently, validation of the particle tracing model is done against the experimental data of Zhang et al.[46] for inertial focusing in serpentine microchannel. Zhang et al. employed a serpentine microchannel with 15 loops and a cross-section of (200 $\mu m$ × 40 $\mu m$). Thus, a channel with AR = 0.2 is employed for particle focusing, having one inlet and one outlet. Polystyrene beads of diameter 9.9 $\mu m$ were used in a fluid with a kinematic viscosity of $6.67 \times 10^{-7} \, m^2/s$ and inlet velocity $1.1 \, m/s$ which corresponds to a channel Reynolds number ($Re_c$) of 110. Under such flow conditions, particles while moving through the microchannel attained a certain streak width within it, in each loop. Under the combined effect of the Dean drag force, inertial lift force, and centrifugal force the particles achieved their corresponding equilibrium positions inside the channel loops. This streak width is plotted against the zigzag loop number, in Fig. 5(b). The results of the streak width plot obtained from the paper by Zhang et al.[46] are compared with the predicted results of the present CFD simulation, and the match was found to be reasonably good. Thus, the numerical model is validated and hence both the particle tracing and laminar flow model can be used subsequently for the actual study.

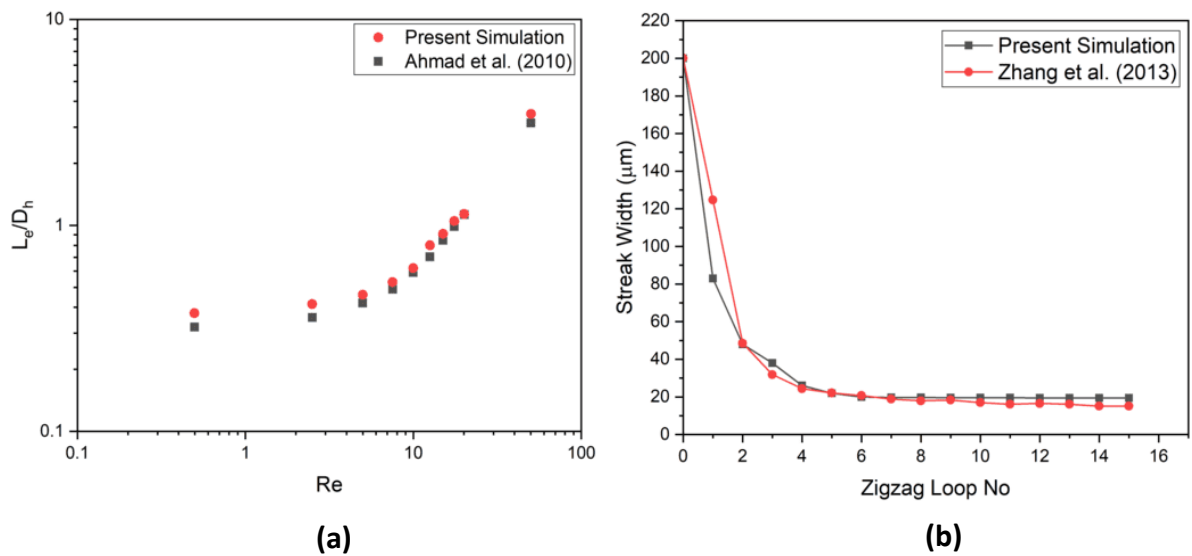

**Figure 5:** (a) Validation of Streak width vs Loop number plot for particle motion within each loop of the serpentine microchannel with literature data. (b) Validation of non-dimensional entrance length vs Re plot with literature data.

## 4. <u>RESULTS & DISCUSSION:</u>

The phenomenon of particle focusing and sorting, under the influence of different forces inside a symmetric serpentine microchannels has been discussed in this section. During the motion of microparticles within a microchannel, initially, they remain randomly distributed through the inlet region, but as they propagate through the microchannel, they tend to accumulate along a single streak in the central part of the microchannel. This phenomenon of migration of microparticles within a channel under the influence of various forces, into a single focused streak is called focusing[14]. For the present study, a symmetric serpentine microchannel has been utilized, as particle focusing can be achieved within a much shorter length, than a straight microchannel with outlet bifurcation[30]. Thus, curvature-induced processes in serpentine microchannels provide a much shorter footprint, which in turn facilitates miniaturization. However, a comparison between serpentine microchannels and straight-channel that takes into account the inertial lift forces is lacking in literature. Figs. 6(a) and 6(b), represent one such comparison between the chosen 30-looped serpentine microchannel and a straight channel of equal length, both having a width of 200 μm. For this comparison n = 1.2 (a/D$_h$> 0.07) and a flow rate of 460 μL/min is chosen.

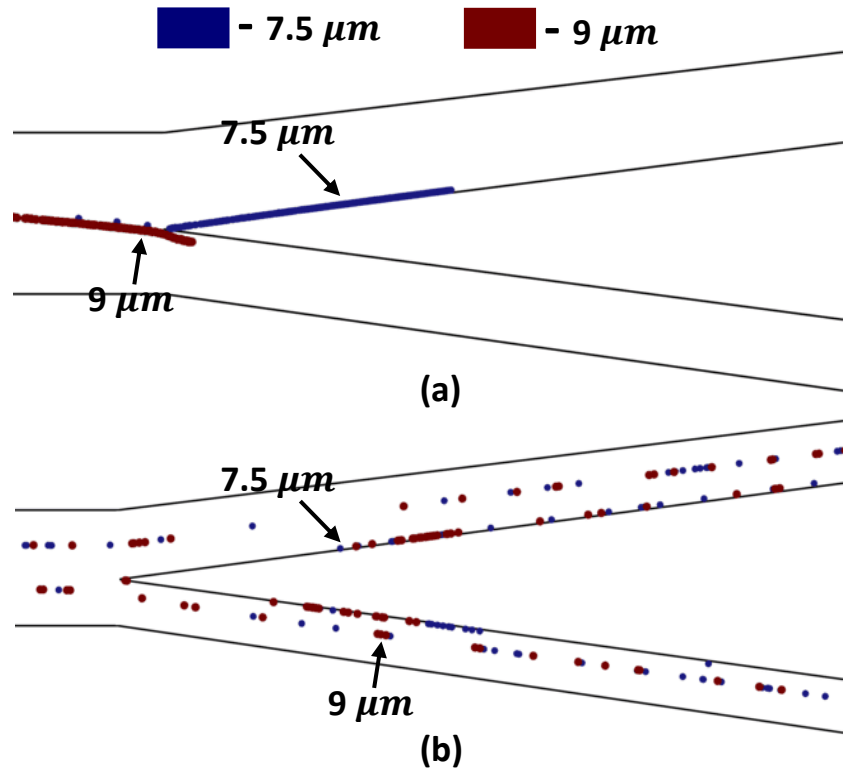

**Figure 6:** Particle distribution at outlet bifurcation for (a) 30-looped symmetric serpentine microchannel and (b) straight microchannel of equal channel length.

### 4.1. Mechanism of particle sorting and contribution of different forces:

A distinctly focused particle streak can be observed from Fig. 6(a), at the bifurcated channel outlet for the chosen 30 looped serpentine channel. On the contrary, no distinct focusing is observed from Fig. 6(b), at the outlet for the straight channel having equal length. Though in both cases the particles traverse the same channel length, focusing is only achieved in the serpentine channel. A larger channel length is required for the straight channel for desirable particle focusing. This is mainly because the centrifugal force, at the serpentine bends, aids the particle-focusing phenomenon. Apart from the inertial lift and drag force acting on the particles, in both cases, the centrifugal force at the serpentine bends, also facilitates the particle streak formation, which ultimately aids in particle focusing and separation at the outlet.

During the fluid motion, within the serpentine microchannel, the fluid streamlines get constricted at each serpentine bend, which leads to higher flow velocity in those regions. The streamlines and velocity contour are depicted in Figs.7(a) and 7(b). Thus, at the bends, the particles get accelerated, due to higher relative fluid velocity. The particles are released with zero velocity, after the steady fluid flow is reached. It is also evident from Eq. 8, that all the particles accelerate due to the drag[64], as the initial particle velocity is lesser than the fluid velocity at the inlet. Since the acceleration is proportional to $a^{-2}$, the smaller particles having lesser radius (a/2) experience more acceleration than the larger ones.

This has been depicted schematically in Figs 7 (c) and (d). For ease of representation, only two particles are considered. As can be seen in Figs. 7(c) and 7(d), the smaller particles pass earlier than the larger particles through the 15th serpentine loop. The smaller particle enters the 15th loop at 0.0413s, and while maneuvering its path away from the outer channel wall, leaves the loop at 0.0428s. Similarly, the larger particle, enters the 15th loop at a later instant of 0.0519s It passes closer to the outer wall and leaves the loop at 0.0541s. In this way the smaller particles move ahead of the larger ones after each bend, and this in turn gets amplified after passing through several loops. Hence, as a result after a certain number of loops, two distinct focused streaks are obtained, where the smaller particle streak moves ahead within the serpentine channel, while the larger particle streak lags. The detailed mechanism of particle focusing and streak formation under the action of different forces within the serpentine channel has been discussed later in this section. Thus, ultimately the smaller particles reach the bifurcated outlet of the serpentine channel earlier, than the larger ones, as shown in Fig. 7(e).

On the other hand, as the particles within a straight channel only move under the action of drag and inertial lift forces, they require a much larger channel length to form any such focused particle streaks[30,31,65]. For the chosen n = 1.2, and flow rate 460 µL/min, the inertial lift (~9 pN) and drag (~35 pN) on particles for a straight channel is comparatively orders of magnitude lesser than inertial lift (~430 pN) and drag (~930 pN) forces at serpentine channel bend. This increased drag and inertial lift forces in the serpentine channels, is mainly due to the accelerating fluid at the serpentine bends. Furthermore, as there are no accelerating bends in a straight channel, the fluid velocity is also less than serpentine channel, as seen in Fig. 7(f). Thus, it takes a longer time and channel length to attain a focused particle streak for the straight microchannel.

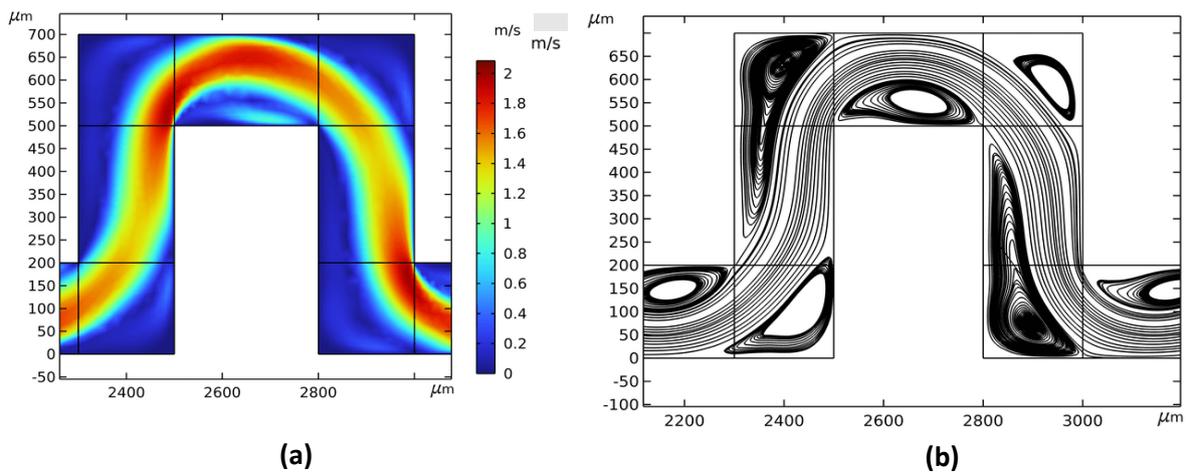

(a)  (b)

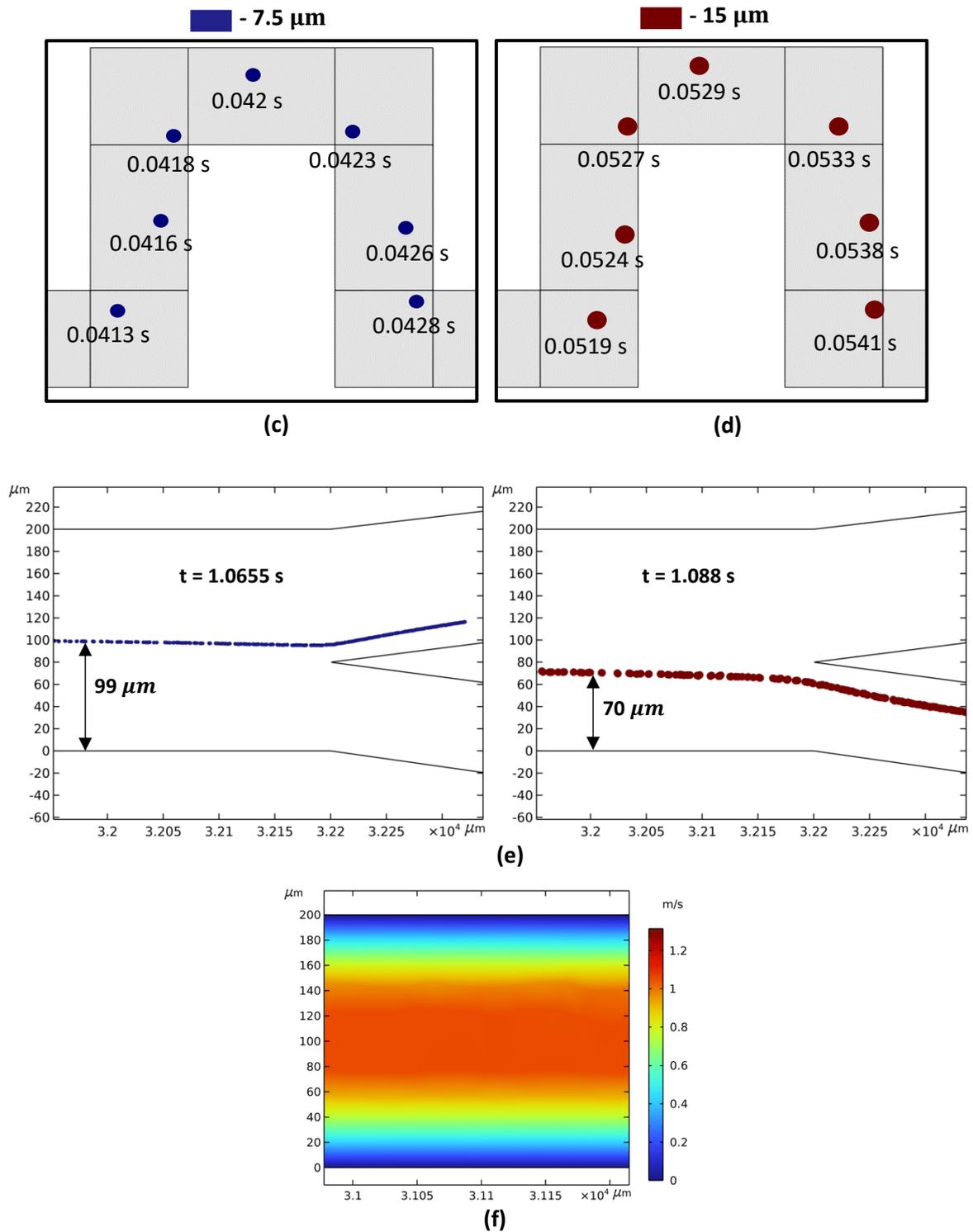

**Figure 7:** (a) Velocity contour and (b) streamline for a serpentine loop. (c), (d) Single particle distribution within the 15$^{th}$ loop, for n = 2, at different time steps. (e) Formation of distinct particle streaks for n = 2, under an appropriate flow rate of 250 µL/min at the outlet bifurcation region of the channel (f) Velocity contour at the middle of the straight channel having the equivalent length of 30 looped serpentine microchannels.

The inertial lift (since $a/D_h > 0.07$), drag and centrifugal forces acting on the particles are the major contributors to this particle focusing and sorting phenomenon. The inertial lift force mainly arises due to the difference in velocity gradient near the wall and central line. Particles near the centerline undergo a lesser relative velocity and this developed shear gradient within the channel pushes the particles toward the walls. Again, for the particles closer to the wall, due to the constricted streamlines between the particles and the wall side, an increased pressure is developed close to the wall, which pushes the particles away from the wall[14,31]. Along with this, bends present in the current microchannel also introduce curvature which leads to the generation of centrifugal forces. This is again one of the major contributors to particle focusing, as seen from literature[44,66]. The expression for centrifugal force is given from the works of Mach et al.[42] and Lim et al.[67] by

$$F_{\text{Cent}} = (\rho_p - \rho_f)\pi a^3 v_{\text{pt}}^2/6r \tag{17}$$

Where, $v_{\text{pt}}$ is the tangential (streamwise) particle velocity and $r$ is the radius of curvature of particle orbit. During the particle motion along the serpentine bends the centrifugal force always acts outward, i.e., from the inner corner to the outer corner of the bend. Depending upon the particle radial velocity and particle position within the channel this centrifugal motion is either opposed or aided by drag and inertial lift forces. Initially while maneuvering a bend, the drag and centrifugal forces act in the same direction, resulting in particle acceleration around the bend. As soon as the particle radial velocity ($v_{pr}$) exceeds the fluid radial velocity ($v_{fr}$) the viscous drag reverses its direction and opposes the centrifugal force. On the other hand, along with this, the inertial lift force also acts on the particles, whose direction is dependent upon the particle position within the channel. For the particles passing along a bend near the channel centreline, inertial lift force pushes the particles toward the channel wall. On the other hand, the particles negotiating the serpentine bend, while being close to the wall, are pushed away, toward the centreline. This phenomenon has been depicted through an illustration in Fig. 8(a). Thus, the trajectory of the particles is dictated by the equilibrium of these forces, acting on them, which ultimately focuses them in a single streak within the channel, after passing through a certain number of loops. Different-sized particles, experience the combination of these forces, at different magnitudes, thus resulting in distinct focused particle streaks for each type of particle. The focused particle streaks of each type attain a constant distance between them, after passing through a certain number of loops. For a particular section of each serpentine loop, the particles get focused to a particular location after a certain focusing distance. This is evidenced by a constant particle distance from the outer wall of the loops. These particle distances from the wall have been determined at the middle of the 1st, 5th, 10th, and 15th serpentine loops, as depicted in Figs. 8(b). As there is a lag between the particles, hence for representation only two particles are considered, and their positions were superimposed to determine their distance from the outer wall. Furthermore, this variation is also plotted for different loop numbers for n = 2, as shown in Fig. 8(c). Irrespective of the particle trajectory in various loops, the particles of different diameters are separated by a constant distance between them. This constant gap ultimately facilitates the separation at the bifurcated outlet. Furthermore, for different diameter ratios, varying numbers of loops are required to effectively form a distinct focused streak. The minimum number of loops required for efficient particle focusing has also been discussed in section 4.2. This formation of the distinctly separated focused particle streaks, for different particle sizes, helps in particle separation at the bifurcated channel outlet, as already represented in Fig. 7(e). Sorting efficiency is enhanced, as the distance between the focused streaks increases. Furthermore, the approximate value of these forces, acting on a particle at a particular bend, for n = 1.2, has been shown in Fig. 8(d). The inertial lift and centrifugal forces are of similar order of magnitude with the drag force. This again supports the fact that inertial forces are also significant for the present study. For particles of $a/D_h > 0.07$, the inertial lift forces, also become significant, as suggested by works of Di Carlo[31,39]. Further discussion regarding the dominance of the inertial lift force, in the present study, has been included in appendix A.2 section.

Hence the combined effect of these three forces causes a specific-sized particle to migrate in the lateral direction and get arranged in a specific streak. The centrifugal force generated due to bends also reduces the total length of the particle sorter by aiding in the focusing phenomenon. However, there is a chance of generation of secondary flows near the channel bends. It is verified that the magnitude of secondary flow is insignificant in the present case which is described in appendix A.3 section.

In the present study, the separation efficiency of small particles has been defined as the ratio of number of collected small particles in the upper outlet to input small particles, and the separation efficiency of larger particles has been defined as the ratio of the number of collected large particles in the lower outlet to input large particle[45]. Furthermore, throughput is defined as the ratio of total number of particles at the outlet to the total input particle[45].

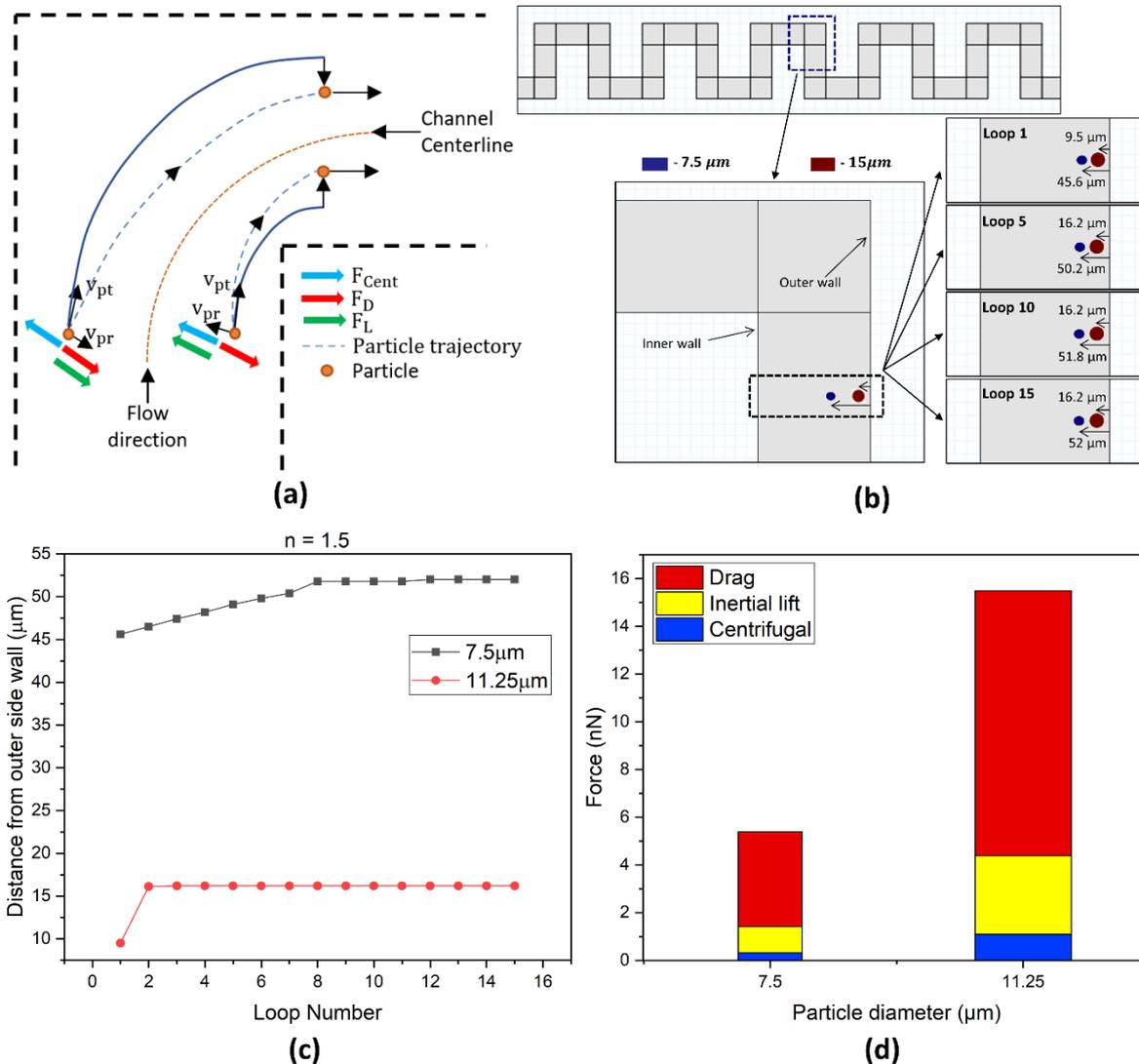

**Figure 8:** (a) Illustration of the particle focusing mechanism (b) Particle distance from the outer channel wall for 1st, 5th, 10th, and 15th loop. (c) Variation of particle distance from the outer wall with serpentine loop number. (d) Chart for the magnitude of different forces acting on 7.5 μm and 9 μm particles.

Thus, it is observed that forces like Stokes drag, inertial lift and centrifugal force combined indeed play a vital role in inertial particle sorting through the present serpentine microchannel. Apart from these forces, other forces like virtual mass force, Archimedes force and Basset force may also sometimes important in the context of particulate flow in inertial microfluidics. But those forces have been neglected in the present study owing to their negligible contribution the particle trajectories, which has been further discussed in appendix, A.5.

### 4.2.1. Effect of channel Reynolds number on particle sorting:

From the earlier literature review, it's been known that inertial migration becomes dominant when $a/D_h > 0.07$ and $Re_p \sim 1$ [14,31,39]. From the works of Di Carlo[31], it is evident that inertial lift forces, i.e., the shear gradient lift, and wall-induced lift force, both play a very vital role in particle focusing in a microchannel. The particle and channel Reynolds number for different diameter ratio combinations, corresponding to varying flow rates, with the microchannel is shown in Fig. 9.

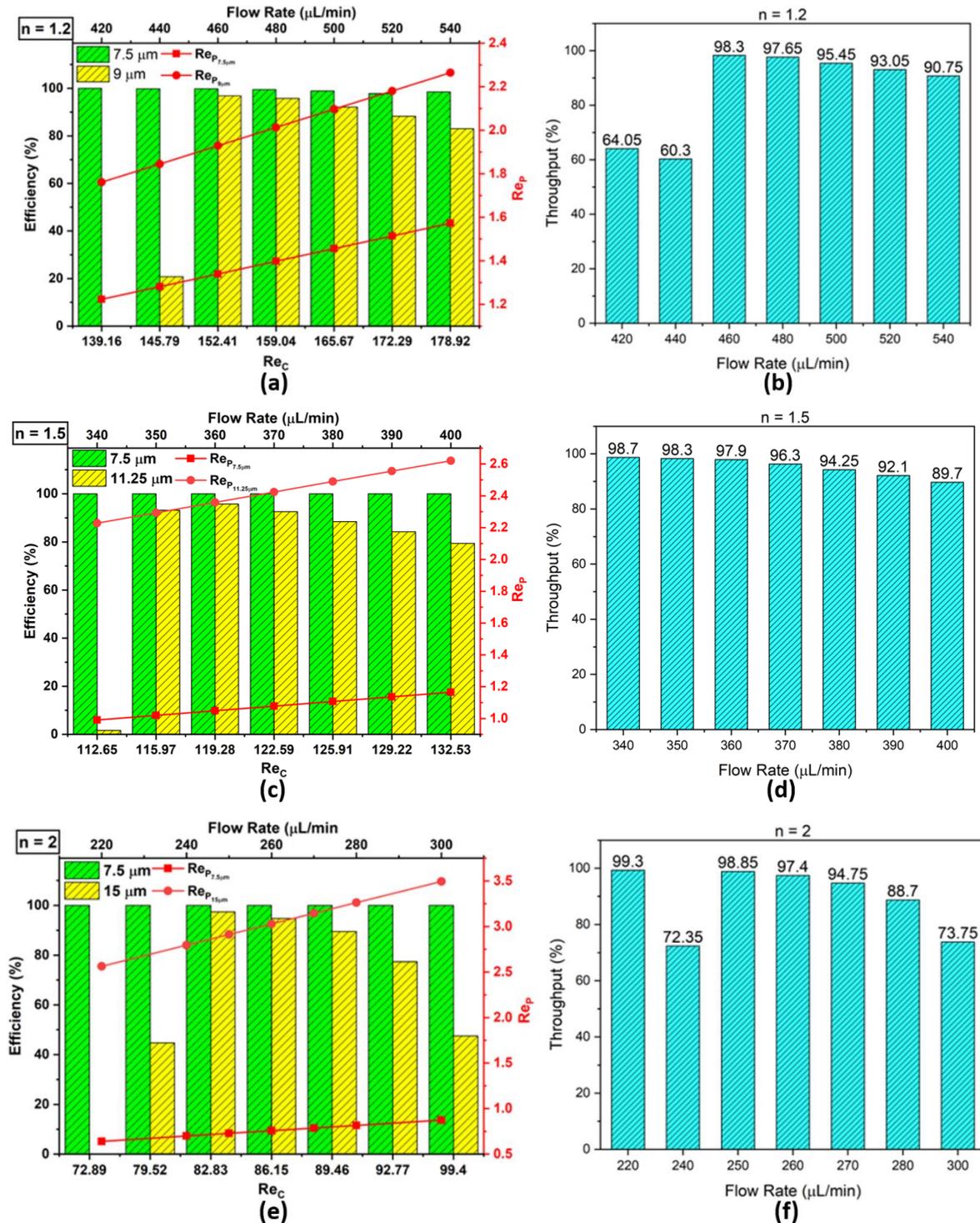

**Figure 9:** (a), (c), (e) Separation efficiency (%), particle Reynolds number and channel Reynolds number; (b), (d), (f) Variation of throughput (%) at different flow rates for diameter ratios 1.2, 1.5, and 2.

Here the variational effect of flow rate on separation efficiency and particle throughput has been discussed. Only those flow rates have been represented in Fig. 9, within which significant changes in sorting efficiency and throughput are observed, corresponding to each diameter ratio.

In this study a channel of width 200 $\mu m$ was chosen, considering three diameter ratios of particle combinations, 1.2, 1.5, and 2, where the minimum particle size is 7.5 $\mu m$. Thus, here, mainly four different-sized particles are considered, viz. 7.5 $\mu m$ ($a_{7.5}/D_h = 0.09$), 9 $\mu m$ ($a_9/D_h = 0.11$), 11.25 $\mu m$ ($a_{11.25}/D_h = 0.14$), 15 $\mu m$ ($a_{15}/D_h = 0.19$)

It is observed that among the different diameter ratios of particles, they all tend to get focused within a focusing streak when $Re_p \sim 1$. From Fig. 9, it can be observed that for a particular hydraulic channel diameter, the particles combination with larger n tend to get sorted at lower flow rates, whereas for particles of smaller n, tend to get sorted at higher flow rates. This may be accounted for because larger particles have more inertial effects than smaller particles for the same flow rate. Hence for larger particles $Re_p \sim 1$ at lower flow rates, whereas for smaller particles it is achieved at higher flow rates.

One thing to be mentioned here is that for 7.5 $\mu m$ particles within the flow rate range of 200-300 µL/min ($Re_c$ = 66.27 - 99.4) have $Re_p < \sim 1$. But still, they attained a focusing streak within the channel. This is because the centrifugal force acting on the particle compensated for the inertial force deficit on the particle. Hence under the combined effect of inertia and centrifugal force focusing is attained by the 7.5 $\mu m$ particles.

It can be observed from the Fig. 9, for a particular diameter ratio, with an increase in the flow rate $Q$ the particle Reynolds number ($Re_p$) keeps on increasing. As the $Re_p$ increases the inertial migration effects keeps on dominating the particle behavior inside the microchannel. Thus, the particle migration within the microchannel and their behavior are dominated by the inertial forces acting on them throughout the channel length and centrifugal force acting on them mainly at the serpentine bends. Consequently, varying-sized particles experience a combined effect of these two forces by varying magnitudes, leading to the formation of differential focused streaks in the central region of the microchannel. Thus, the particles can predominately migrate across the streamlines, perpendicular to the direction of primary fluid flow, and attain equilibrium focusing positions. The lighter and smaller particles are focused on the upper side and alternately the bigger heavier particles are focused below the centre line. Mainly this differential focusing streaks at different heights of the channel facilitates the separation of the particles by the bifurcated channel outlet. Simulations were performed under different flow rates to determine the maximum throughput and sorting efficiency for different particle diameter ratios. It can be seen from Figs. 9(a), 9(c), and 9(e) for diameter ratios of 1.2, 1.5, and 2, a particle separation efficiency of 96.8 %, 95.8 %, and 97.4 % has been observed, under an appropriate flow rate of 460 µL/min ($Re_c$ = 152.41), 360 µL/min ($Re_c$ = 119.28) and 250 µL/min ($Re_c$ = 82.83) respectively.

It is noteworthy that, there is a cut-off channel Reynolds number for each of the diameter ratio combinations below which the separation efficiency is very low, where hardly any separation occurs. For instance, from Fig. 9(a), in case of diameter ratio of 1.2, for a flow rate of 440 µL/min ($Re_c$ = 145.79), the separation efficiency is as low as 20.8 %. Similarly, from Figs 9(c) and 9(e), at a corresponding flow rate of 340 µL/min ($Re_c$ = 112.65) and 240 µL/min ($Re_c$ = 79.52), the separation efficiency is 1.7 % and 44.7 % for diameter ratios of 1.5 and 2, respectively. This can be accounted, because of the combined low inertial and centrifugal forces at a flow rate below the threshold, though focusing is achieved in the microchannel, but these forces were not sufficient enough to distinctly produce differential focusing streaks, for the particles to get sorted at the bifurcated outlet. As the $Re_c$ increases the throughput increases along with the increase in the separation efficiency, up to a certain limit, as depicted in Fig. 9.

Again, with increasing $Re_c$, a decreasing trend is observed from Fig. 9, after a certain limit, for both throughput and separation efficiency. For instance, from Figs. 9(a) and 9(b), in case of diameter ratio 1.2, the separation efficiency of bigger particles drops down from 96.8 % to 83 % within the microchannel, and the throughput also drops to 90.75 % at a flow rate of 540 µL/min ($Re_c$ = 178.92). From Figs. 9(c) and 9(d), for diameter ratio 1.5, the sorting efficiency and throughput also drop down to 79.4 % and 89.7 % respectively, at a flow rate of 400 µL/min ($Re_c$ = 135.53). Similarly, from Figs. 9(e) and 9(f), for diameter ratio 2, the sorting efficiency and throughput drop down to 47.5 % and 73.75 % respectively, under a flow rate of 300 µL/min ($Re_c$ = 99.4). The reason behind sudden drop in sorting efficiency on high flow rates, is related to Stokes number and related particle entrapment, which is discussed in the next section.

### 4.2.2. Effect of Stokes number on particle sorting:

The stokes number (*Stk*) plays a significant role in particle trajectory. Particles with lower Stokes number, tends to follow the fluid streamlines around a bend, where as for larger values of Stk, the particle tends to get deraigned from fluid streamlines and many tines get stuck inside the micro vortices within the channel and this will be discussed in detail in the following section. The effect of *Stk* on sorting efficiency of the bigger particle for each particle diameter duo was investigated extensively and has been plotted in Fig. 10. Here, Fig. 10(a) shows the variation of sorting efficiency with Stk, for n = 1.2, 1.5 and 2 (base particle diameter kept constant to that of 7.5 µm). It is observed that, with increasing Stk, which is linearly dependent on $Re_c$, the sorting efficiency of the big particle initially increases, due to the increased drag and lift force from Eqs. 8, 9 and 10. Furthermore, the plot clearly depicts a decreasing trend in sorting efficiency, as the Stk increases (~(O)1) along with the increasing flow rate for each particle duo combination, which is mainly accounted due to particle entrapment. Additionally, the effect of variation of base particle diameter ($a_{base}$) has also been analyzed. Since all the other diameter ratios also show similar trend, here a representative case for n = 1.2 is shown in Fig. 10(b) for $a_{base}$ = 7.5 µm, 15 µm and 22.5 µm. Similar trend in particle sorting efficiency is observed with the variation of *Stk* for various base particle diameters.

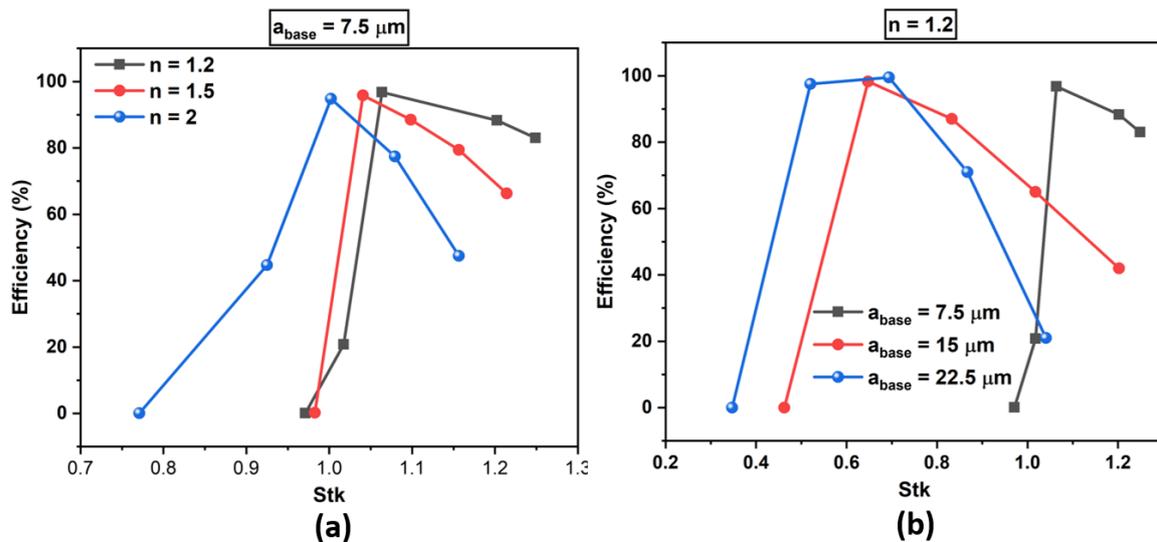

**Figure 10:** (a) Variation of sorting efficiency with Stk, for n = 1.2, 1.5, 2, with base diameter 7.5 µm. (b) Variation of sorting efficiency with Stk, with $a_{base}$ = 7.5 µm, 15 µm and 22.5 µm for n = 1.2.

The decrease in efficiency with the increase in *Stk* may be either due to particle entrapment within the micro-vortices or because of the mixing effects of secondary Dean drag within the microchannel.

However, in the present study, for the chosen microchannel dimensions and flow rate ranges, mixing effects are not dominant here (see appendix: A.3). It is observed that, as the fluid enters the channel with high inertia for larger values of $Re_c$, many particles entering the fluid gets stuck inside the loop, along the channel walls, where the boundary layer velocity is ~ 0. Due to the increase in fluid flow rate, the *Stk* of the particle as in Eq. 7, also increases. So, the particles tend to deviate from the fluid streamlines. Thus, instead of following the mainstream fluid flow, they deviate and get stuck along the laminar boundary layer at the microchannel walls. Again, it's also seen, that particles also get stuck inside the micro-vortices formed at the serpentine bends. At higher flow rates, near the serpentine bends, the larger particles couldn't follow the natural streamline curvature, instead, they get entrapped within the micro-vortices and remain there. This in turn leads to less throughput at the outlets and the separation efficiency also gets significantly affected. So, the decrease in separation efficiency and throughput after a certain upper threshold flow rate is accounted mainly due to entrapment of particles within the channel. This entrapping of microparticles within the channel leads to clogging, which can be a scope of further investigation. Figure 10 shows the particle trajectories superimposed with velocity vectors inside loop 1 of the serpentine channel, for n = 2, at a flow rate of 340 μL/min ($Re_c$ = 112.65).

From Fig. 11(a), it is seen that, on the left side of the 1st loop the smaller particles tend to follow the streamline motion along a bend, while the larger particle having greater inertia gets deflected towards the outer channel wall. Again Fig. 11(b) and 11(c) show the entrapping motion of the larger particle within the micro vortices, formed at the right corner of the 1st loop. It also shows, many larger particles getting stuck on the channel side wall, within the laminar boundary layer region, while other smaller particles manoeuvre the bend following the streamline motion.

Many times, such a scenario may arise, where sorting efficiency is very poor, even though collected sample purity is high. Under appropriate flow conditions, only those cases are considered for particle collection, where a high percentage of sorting efficiency, throughput, and sample purity are obtained. Again, from a perspective of medical diagnosis, purity of the sample collected at the outlets of the microchannel, is also an important parameter contributing to the usability of the serpentine microchannel. The purity of the particles is plotted in Fig. 12, corresponding to each outlet. Here the purity of a particular sample is defined as the ratio of the number of target particles at a particular outlet to the number of total particles at that outlet[45]. Thus, depending on. the combined results of sorting efficiency, throughput, and purity, a particular range of flow rate is appropriate for effective sorting of varying-sized particles within the microchannel.

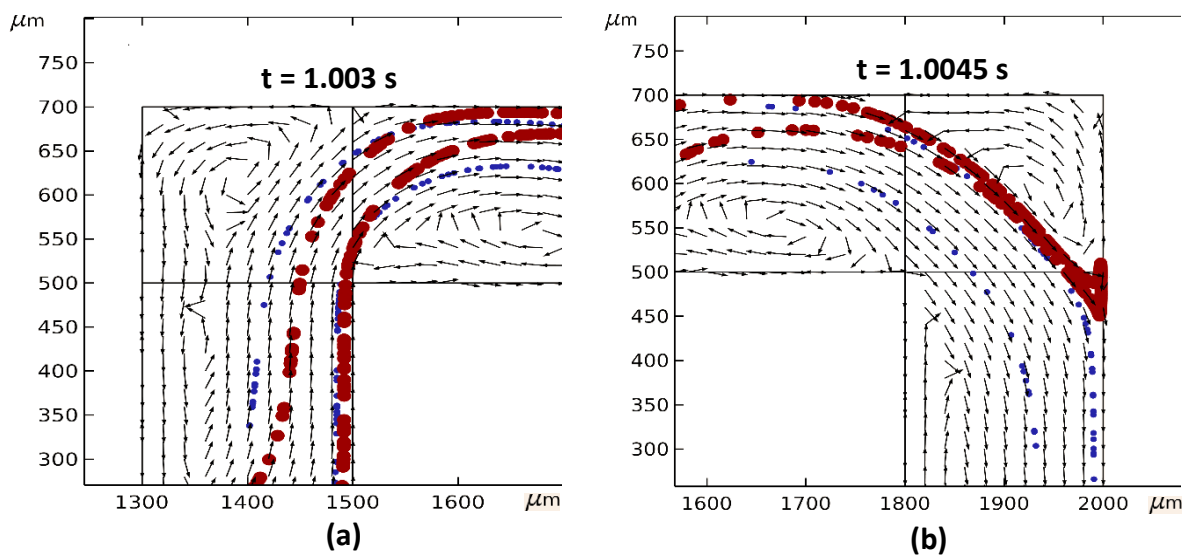

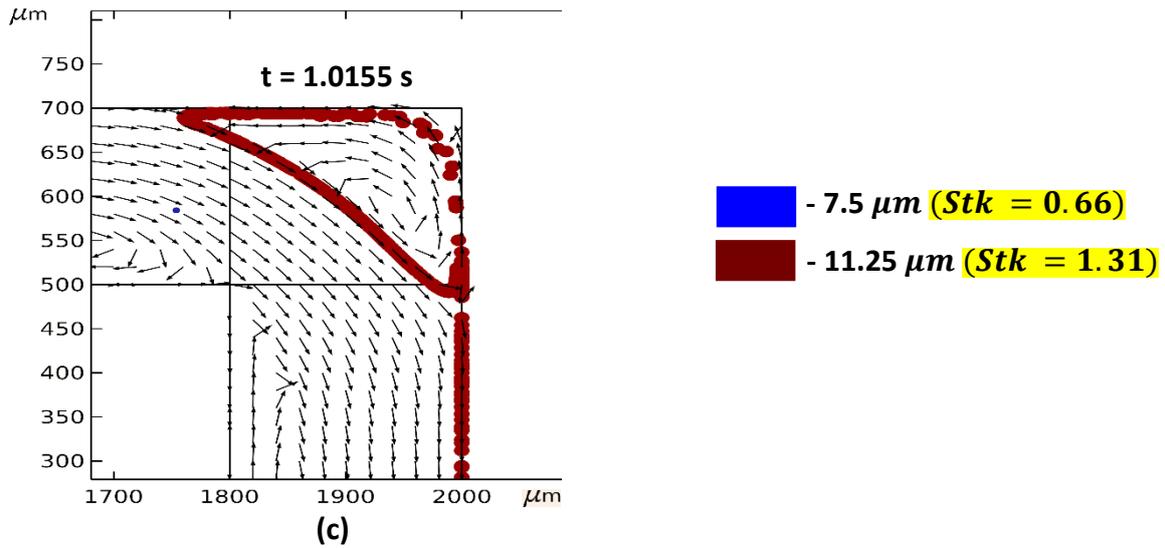

**Figure 11:** Particle motion and velocity vector on (a) left side and (b) right side of the 1st serpentine loop (c) Entrapped particles in micro vortices on right side of the 1st serpentine loop.

It can be seen from Figs. 12(a), 12(b), and 12(c) that, for diameter ratios 1.2, 1.5, and 2, they have very less sample purity, at corresponding low flow rates of 420 µL/min, 340 µL/min and 200 µL/min. As discussed earlier, this is due to the fact, that at these flow rates, the combined effect of inertial and centrifugal forces, is inadequate to form distinct focused particle streaks, for the particles of the given diameter ratio. Hence further highlighting the need for an appropriate range of flow rate for achieving high sorting efficiency, throughput and purity.

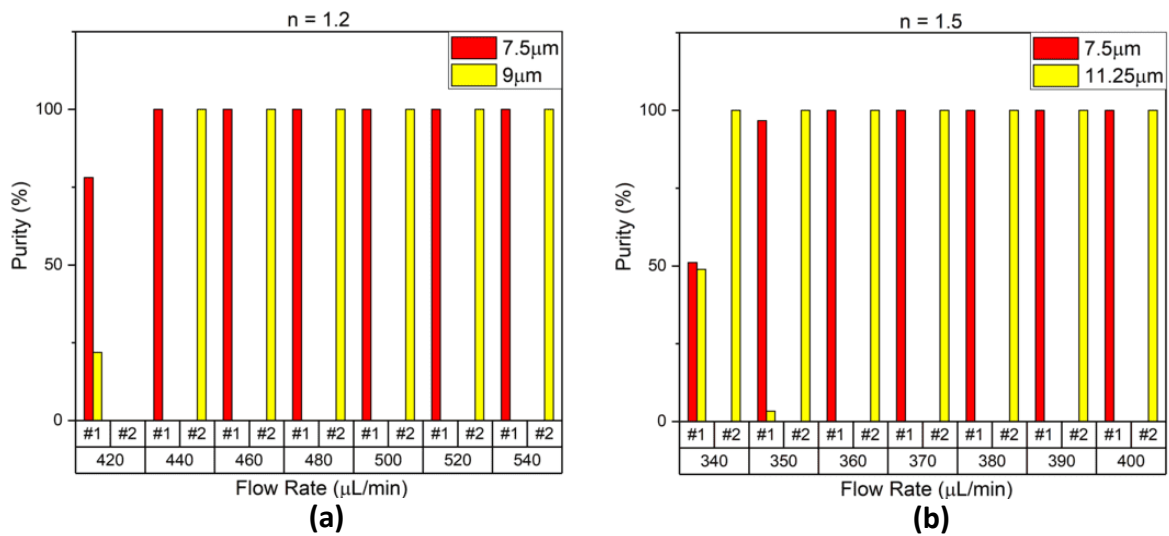

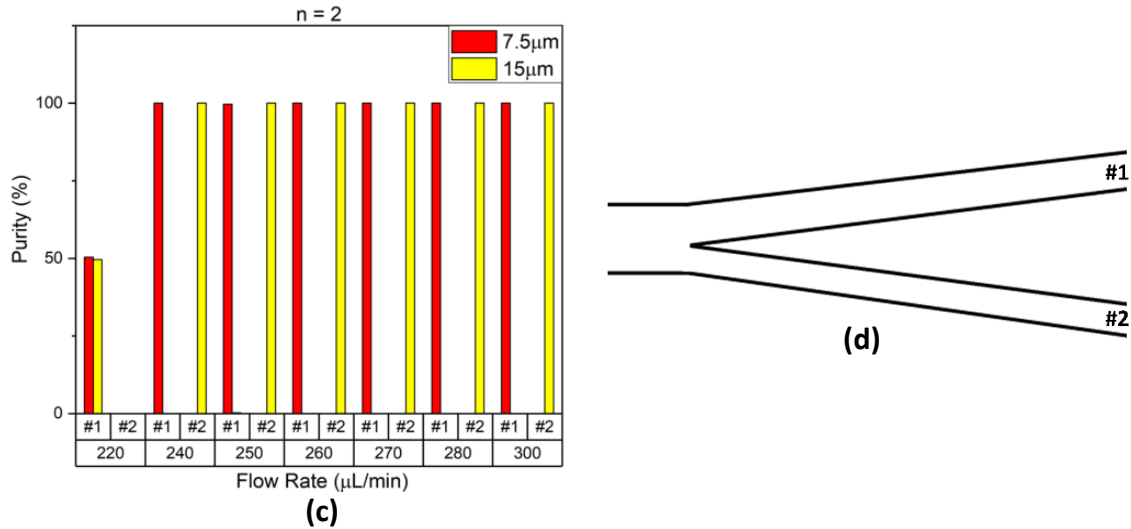

**Figure 12:** Variation of purity (%) at different flow rates for diameter ratios (a) 1.2 (b) 1.5 and (c) 2 (d) "#1" and "#2" indicates the upper and lower outlet of the microchannel respectively.

### 4.2.3. <u>Effect of density ratio in sorting:</u>

The density ratio (α) is defined as the ratio between the particle and fluid density within the channel. It has been found that density ratio (α) also plays a vital role in the focusing length and sorting of microparticles within the channel. Initially for the chosen fluid and particle density, α = 1.12. Consequently, keeping the particle density unaltered, the fluid density is altered and subsequently 5 values of α are chosen, viz. 0.82, 0.92, 1.02, 1.12 and 1.22. Simulations were performed, at a constant flow rate of 360 μL/min for n = 1.5 in a 8 loop serpentine channel, to analyze the correlation between sorting efficiency and density ratio. As shown in Fig. 13(a), with decrease in the density ratio (α), the sorting efficiency also drop significantly, leading to no particle separation at the bifurcated channel outlet.

This is accounted due to the insufficient forces acting on the particles in case of lower density ratio. With increasing α, the overall lift, drag and centrifugal forces on the particles also increases. It can be observed from Fig. 13(a), with increasing α, the channel Reynolds number ($Re_c$) decreases, which in turn leads to higher Stokes drag and lift force, as evident from Eq. 14 & 15. This combined hydrodynamic forces on the particles for higher α (1.02 onwards), forms focused particle streaks for two particles at a sufficient distance from each other. It is noteworthy here, that in all the cases of α, focused particle streaks are formed. But for α = 0.82, 0.92, they are so closely formed that all the particles enter the upper bifurcation, which results in poor sorting efficiency. Thus, for higher α, due to the larger separation of the focused particle streaks, significant sorting efficiency for both particles are obtained. Some approximate values of these forces acting on the particles at a particular bend for two α values 0.82 and 1.12 have been shown in Fig. 13(b) and 13(c), for n =1.5.

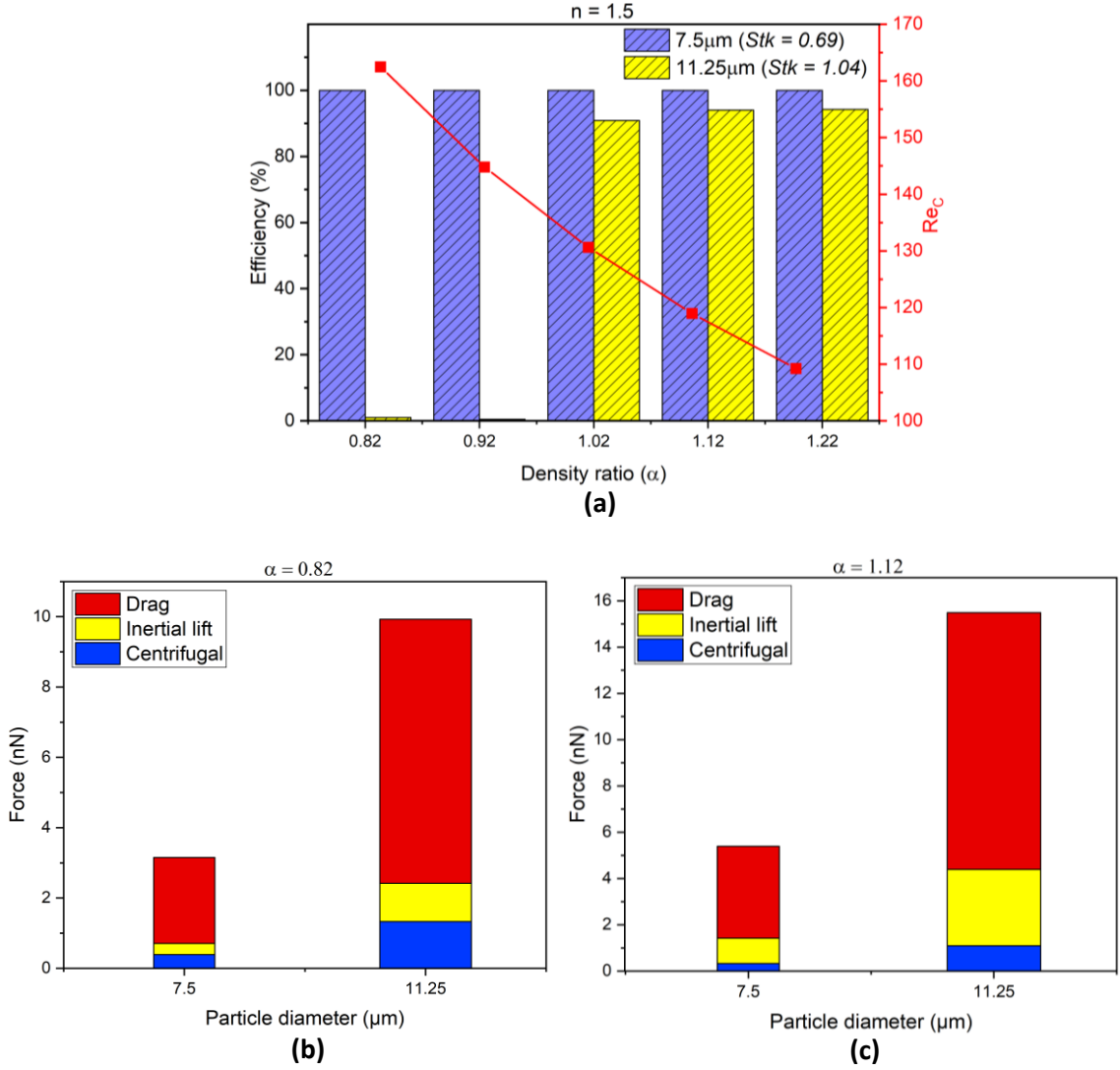

**Figure 13:** (a) Sorting efficiency (%) and channel Reynolds number ($Re_c$) variation with density ratio for n = 1.5. Drag, inertial, and centrifugal force values for (b) α = 0.82 and (c) α = 1.12.

### 4.3. Determination of the minimum number of loops for efficient particle focusing:

Till now the various aspects effecting the particle sorting in the serpentine channels has been explored. Another, important aspect that needs to be determined from an engineering perspective is the minimum number of loops required for achieving particle sorting under certain flow parameters. Herein, a scaling analysis based on particle velocity considerations has been implemented to predict the required number of loops based on the flow parameters. Besides the particle tangential velocity, which effects the centrifugal force (equation 17) in serpentine microchannels; the particle radial velocity is equally important for particle sorting. There are mainly two components dictating the particle radial velocity: radial fluid velocity $v_{fr}$ and the particle centrifugal force. So as suggested by Zhang et al.[46], a parameter of relative radial particle velocity, $v_{prr}$ is formulated. It is the relative velocity of the particle with respect to the fluid in the radial direction. Thus, it indicates the net radial migration of the particles under the effect of centrifugal force.

$$v_{prr} = v_{pr} - v_{fr} = (\rho_p - \rho_f)a^2 v_{pt}^2 / 18r\mu \qquad (18)$$

The ratio of the distance traversed by a particle perpendicular to the streamline to the distance along the streamline is defined as the particle focusing efficiency ($\delta$).

$$\delta \sim \frac{L_r}{L_t} \sim \frac{v_{pr}}{v_{pt}} \sim \frac{v_{fr}}{v_{pt}} + \frac{(\rho_p - \rho_f)a^2 v_{pt}}{18r\mu} \tag{19}$$

Another parameter $\eta_{Cent}$ is defined as the ratio of $v_{prr}$ and $v_{pr}$ to quantify the significance of the effect of centrifugal force in altering the particle motion and aiding in focusing.

$$\eta_{Cent} \sim \frac{1}{1 + \frac{18 v_{fr} r \mu}{(\rho_p - \rho_f) a^2 v_{pt}^2}} \tag{20}$$

Now for effective focusing the particle needs to transversely migrate half of the channel width w, to get focused into the center of the channel, within the minimum focusing length $L_{min}$ of the channel,

$$L_{min} \sim \frac{w/2}{v_{pr}} v_{pt}$$

$$L_{min} \sim \frac{w v_{pt}}{2\left\{v_{fr} + \frac{(\rho_p - \rho_f)a^2 v_{pt}^2}{18r\mu}\right\}} \tag{21}$$

From Bhagat et al.[37,68] the lateral migration velocity of particles in a channel under the influence of inertial lift forces is given by

$$U_L = \frac{\rho_f U_m^2 a^3}{6\pi \mu D_h^2} \tag{22}$$

Taking $U_f = \frac{2}{3} U_m$, (assuming fully developed flow) where $U_f$ and $U_m$ are the average and maximum fluid velocity respectively[31]. So, $U_f = \frac{2\mu Re_c}{3\rho_f D_h}$. By substituting $v_{fr} = U_L$ and $r = w$,

$$\delta \sim \frac{3\rho_f U_f^2 a^3}{U_f(8\pi \mu D_h^2)} + \frac{2(\rho_p - \rho_f)a^2 Re_c \mu}{18r\mu(3\rho_f D_h)}$$

$$\delta \sim \frac{Re_c}{4\pi}\left(\frac{a}{D_h}\right)^3 + \frac{(\rho_p - \rho_f)a^2 Re_c}{27 w \rho_f D_h} \tag{23}$$

$$\eta_{Cent} \sim \frac{1}{1 + 18r\mu \frac{3\rho_f U_f^2 a^3}{(8\pi \mu D_h^2)(\rho_p - \rho_f)a^2 U_f^2}}$$

$$\eta_{Cent} \sim \frac{1}{1 + \frac{27 w a \rho_f}{4\pi(\rho_p - \rho_f) D_h^2}} \tag{24}$$

$$L_{min} \sim \frac{w U_f}{2\left\{\frac{3\rho_f U_f^2 a^3}{8\pi \mu D_h^2} + \frac{(\rho_p - \rho_f)a^2 U_f^2}{18r\mu}\right\}}$$

$$L_{min} = k_{scale} \frac{w D_h}{Re_c\left\{\frac{a^3}{2\pi D_h^2} + \frac{2(\rho_p - \rho_f)a^2}{27 w \rho_f}\right\}} = k_{scale} f(a, Re_c) \tag{25}$$

$$N_{min} = k_{scale} \frac{w D_h}{l Re_c\left\{\frac{a^3}{2\pi D_h^2} + \frac{2(\rho_p - \rho_f)a^2}{27 w \rho_f}\right\}} = k_{scale} f(a, Re_c) \tag{26}$$

$$L_{min} = k_{scale} L_{loop}$$

Here $k_{scale}$ is the scaling factor. Now, nondimensional scaling parameters are defined for $L_{min}$

$$\tilde{L}_{min} = \frac{L_{min}}{D_h}, \quad \tilde{w} = \frac{w}{D_h}$$

Further another dimensionless parameter, confinement ratio is defined as $\xi = \frac{a}{w}$,

$$\tilde{L_{min}} = k_{scale} \frac{1}{\xi Re_c \left\{ \frac{\tilde{a}^2}{2\pi} + \frac{2}{27}(\alpha-1)\xi \right\}} = k_{scale} f(\tilde{a}, Re_c) \tag{27}$$

. So, the minimum length of total number of loops required for focusing is

$$\tilde{L_{min}} = k_{scale} \tilde{L_{loop}} \tag{28}$$

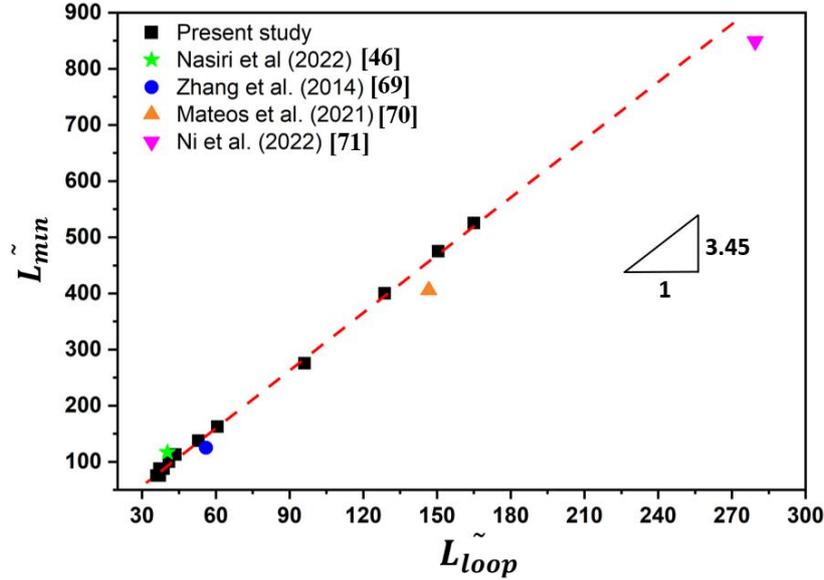

**Figure 14:** Plot showing the linear correlation between the minimum number of loops required for sorting and that predicted by the scaling analysis; plot depicts data both from present study and that available in literature [46,69–71].

This $\tilde{L_{min}}$ as predicted by equation 28 is a linear function of $\tilde{L_{loop}}$; $\tilde{L_{loop}}$ is defined in equations 27-28. The $\tilde{L_{min}}$ obtained from simulations when plotted against $\tilde{L_{loop}}$ falls on a straight line as depicted in Fig.14 above, thus proving the authenticity of equation 28. Again, the close fit of the previous literature data as mentioned in the Fig. 14, further testify the validity of the present scaling analysis along with its wide range of applicability under varying flow conditions and sorting length. Furthermore, the slope of the line as shown in Fig.14 provides the magnitude of the scaling parameter k$_{scale}$. Thus the magnitude of k$_{scale}$ along with equation 28 allows prediction of the minimum number of total loop length required for particle sorting thus providing design guidelines for serpentine microchannels implemented in particle sorting.

Furthermore, to demonstrate the robustness of our proposed scaling laws, we validate Eq. 28 against data from existing literature [45,69–71], as shown in Fig. 14. Both the results from our study and those reported in the literature adhere to the linear trend as described by Eq. 28, underscoring the universality and reliability of our scaling laws.

## 5.  **<u>Conclusions and Outlook</u>**:

Focusing and sorting of particles in a symmetric serpentine microchannel has been presented in the current study considering both the influence of inertial and centrifugal forces only. It has been successfully demonstrated that focusing and efficient sorting of microparticles can be achieved, without the intervention of secondary flow. The key findings of the study have been summarized below.

- Inertial migration in microchannels was shown to be a critical factor in achieving effective particle sorting, particularly when the particle-to-hydraulic diameter ratio ($a/D_h$) exceeds 0.07 and the particle Reynolds number ($Re_p$) approaches 1. Through simulations in a symmetric serpentine microchannel, optimal flow rates were identified for diameter ratios of 1.2, 1.5, and 2, achieving sorting efficiencies of 96.8%, 95.8%, and 97.4% at flow rates of 460 µL/min, 360 µL/min, and 250 µL/min, respectively. Larger particles exhibited inertial focusing at lower flow rates due to stronger inertial effects, while smaller particles required higher flow rates. The combined effects of inertial lift and centrifugal forces enabled smaller particles to achieve focusing even when $Re_p<1$, highlighting the importance of curvature-induced forces. Beyond certain threshold flow rates, sorting efficiency and throughput declined due to particle entrapment within micro-vortices or boundary layers. The particle entrapment detrimentally effects the particle throughput (throughput is defined as the ratio of particles exiting the channel to that leaving the channel), an important parameter in biomedical applications. This entrapment also impacts the purity (purity of a particular sample is defined as the ratio of the number of target particles at a particular outlet to the number of total particles at that outlet) of the particles obtained in the outlet of the serpentine microchannel. Thus degerming the appropriate flow rate is very important for obtaining required sorting efficiency, throughput and purity.

- The density ratio ($\alpha$)—defined as the ratio of particle density to fluid density— plays a critical role on sorting efficiency and focusing behavior of microparticles in serpentine microchannels. Through simulations with α values ranging from 0.82 to 1.22, it was observed that higher density ratios significantly enhance sorting efficiency by amplifying the hydrodynamic forces, including lift, drag, and centrifugal forces, acting on the particles. While focused streaks were formed across all cases, lower α values (0.82 and 0.92) resulted in closely spaced streaks, leading to poor separation at the bifurcated outlet. Conversely, for α values of 1.02 and above, the increased spacing between focused streaks facilitated effective sorting. These findings highlight the importance of optimizing the density ratio to achieve precise particle separation, offering valuable guidance for designing high-performance microfluidic systems tailored to specific particle-fluid combinations for biomedical diagnostics and industrial applications.

- A scaling analysis framework to predict the minimum number of loops required for effective particle sorting in serpentine microchannels is established. By analyzing the relationship between particle velocity components—radial velocity driven by centrifugal forces and tangential velocity—the study identifies key parameters influencing particle focusing, including the hydraulic diameter, channel width, particle size, flow rate, and density ratio. A dimensionless parameter, $\eta_{Cent}$, was introduced to quantify the contribution of centrifugal forces to particle motion, while the minimum focusing length ($L_{min}$) was derived as a function of these parameters. The analysis reveals a linear correlation between the minimum number of loops ($N_{min}$) and the loop length factor ($N_{loop}$), as validated through simulations. This linear relationship, governed by a scaling factor ($k_{scale}$), provides an accurate predictive model for determining the loop count required to achieve sorting under various flow conditions. These insights offer practical design guidelines for optimizing serpentine microchannels, ensuring efficient particle sorting with minimal channel length, thereby supporting the development of compact, cost-effective microfluidic systems for biomedical and industrial applications. Additionally, the robustness of the proposed scaling laws is demonstrated by their consistency with findings from previous studies, which exhibit the same linear trend, underscoring the universality and reliability of the model. These insights offer practical design guidelines for optimizing serpentine microchannels, ensuring efficient particle sorting with minimal channel length, and supporting the development of compact, cost-effective microfluidic systems for biomedical and industrial applications.

By this study, particle focusing can be obtained within a state-of-the-art length, with the significant advantage of high sorting efficiency and throughput. Again, from the scaling analysis, one can get an idea of the required minimum number of loops for a particular diameter ratio combination, which in turn will facilitate design and cost-effective fabrication of the microchannel. In conclusion, this study provides a cost-effective, simple focusing mechanism for microparticles which can potentially be applied for filtration, flow cytometry, bio-cell sorting, and many other biomedical diagnostic applications.

# APPENDIX:

## A.1. Comparison of 2D and 3D Results:

In general 3D simulations of flow, and physics are computationally very expensive. Hence a 2D simulation replicating the results of the actual 3D simulation is desirable. For this purpose, a 3D microchannel geometry was modeled, having the same AR = 0.25 and other dimensions as the 2D model mentioned earlier. The three-dimensional width of the channel is 50 μm. A point probe is considered at the same location as the 2D model, at a distance of 25 μm from each face of the 3D microchannel model, as shown in Fig. 15(a) and 15(b).

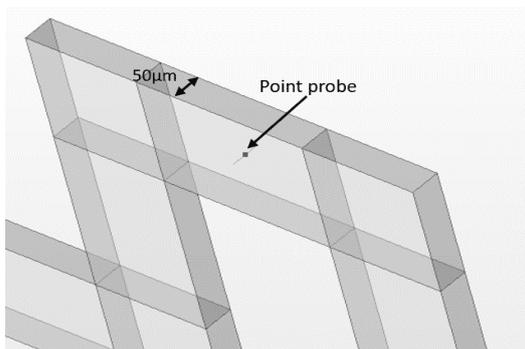

(a)

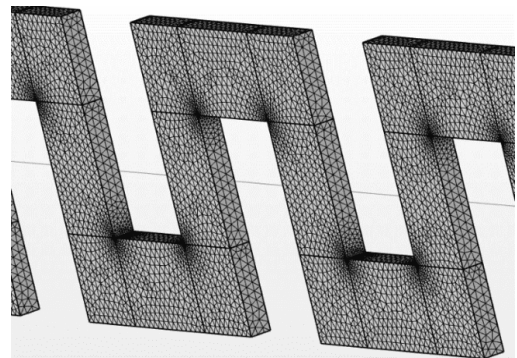

(b)

| Model | Number of mesh elements | Probe velocity magnitude (m/s) | Error (%) |
|---|---|---|---|
| 3D | 1605702 | 0.71663 | |
| 2D | 109394 | 0.70704 | 1.34 |

(c)

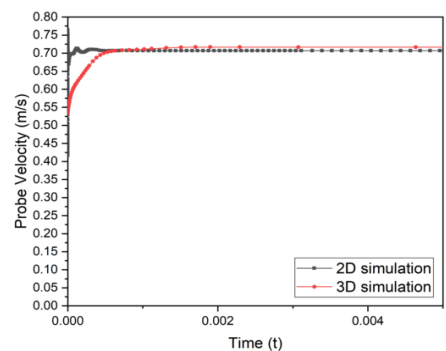

(d)

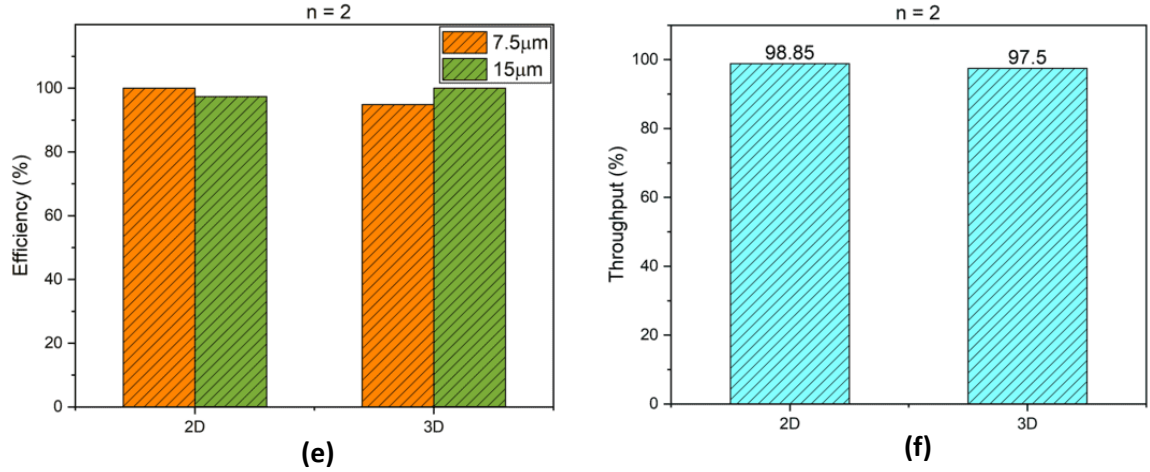

**Figure 15:** (a) Probe point location inside the 3D geometry. (b) Adopted mesh for the 3D model of the microchannel. (c) 2D and 3D mesh geometry and probe velocity comparison. (d) Probe velocity comparison plot for 2D and 3D geometry. (e) 2D vs 3D particle separation efficiency at the outlet for n = 2. (f) Particle throughput comparison for both 2D and 3D.

The probe point velocity of the fluid is determined and compared for both 2D and 3D models, depicted in Fig. 15(c), 15(d) under a flow condition of Q = 260 μL/min, corresponding to $Re_c = 86.147$. The particle distribution and throughput for 2D and 3D models are also compared under the same flow conditions, as shown in Figs. 15(e) and 15(f). From the above analysis, this 2D model compares very closely to the full 3D simulation in terms of both flow profile and particle distribution. Thus, for the present numerical study, a 2D model is used to replicate the actual fluid flow physics and particle motion of a 3D microchannel geometry.

### A.2. Effect of inertial lift force:

Another salient feature of the present study is the dominance of inertial forces, for the chosen channel dimensions and particle combinations. As mentioned earlier, it has been seen from the works of Di Carlo[31] that inertial effects are dominant for $a/D_h > 0.07$. Hence to support this claim, two different simulations were conducted while maintaining a channel Reynolds number of $Re_c = 152.41$. The first simulation used the present model with a 200 μm wide channel, while the second featured a 600 μm wide serpentine channel with 30 loops, as shown in Fig. 16. For the given n = 1.2, the ratio $a/D_h$ is greater than 0.07 in the 200 μm channel and less than 0.07 in the 600 μm channel. Since the Reynolds number is the same in both cases, the drag force acting on the particles remains identical, and due to the similar serpentine geometry, the centrifugal force is also comparable. Consequently, particle separation and focusing are governed solely by the magnitude of the inertial lift force, allowing us to assess its effect on particle focusing.

In the case of the 600 μm serpentine channel, due to the lack of inertial lift forces (~0.4 pN) on the particles, no focusing streaks are observed and hence produce very poor particle sorting efficiency, as seen in Fig. 16(a). On the other hand, in case of 200 μm serpentine channel chosen in the present study, although the particles traversed the same channel length, a focused streak along with high sorting efficiency has been achieved, under identical flow conditions. Hence it is evident, that inertial lift force does play a vital role in the context of the present study. Figure 16(b) compares the sorting efficiency for the 200 μm and 600 μm channels.

Furthermore, another simulation was also performed for n = 1.5, at Re = 119.28 (Q = 360 μL/min) for the 200 μm serpentine channel, in order to analyze the effect of inertial lift. Two cases were simulated, one with the lift force node added and another without lift force node in COMSOL. As in the case of n = 1.2, $a/D_h > 0.07$ for both the particles, hence inertial forces should be significant on the. Subsequently, as observed from Fig. 16(c) with the lift node added, high separation efficiency is achieved (~96%), whereas in the absence of lift force, very poor separation (~2.1%) is obtained.

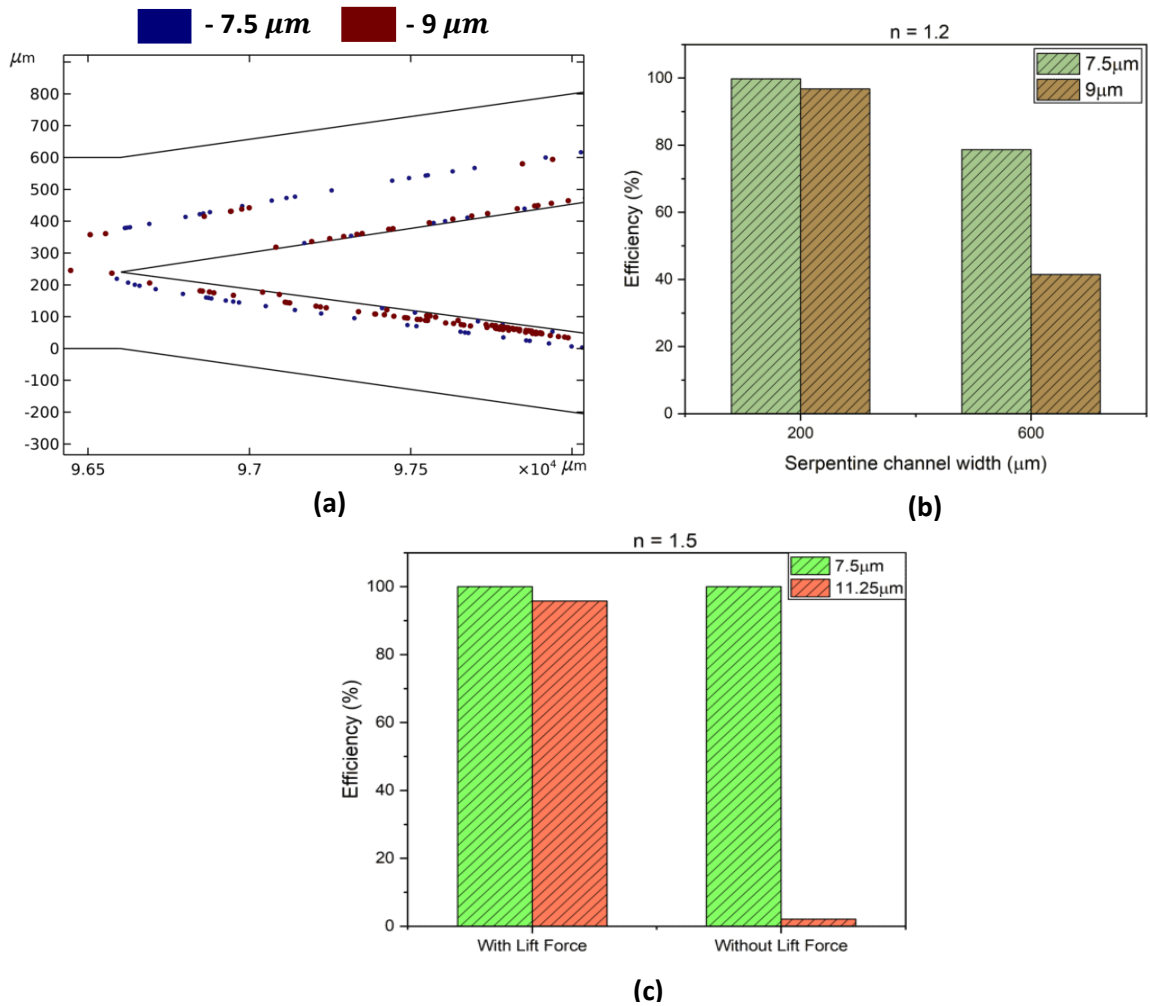

**Figure 16:** (a) Unfocused particle distribution at the 600 μm channel outlet. (b) Sorting efficiency comparison for n = 1.2, in the 200 μm and 600 μm serpentine channel, both having 30 loops. (c) Separation efficiency comparison for n = 1.5, for with and without lift force node.

### A.3. Mixing effects of Secondary flow:

Again, apart from the Inertial lift forces and Centrifugal forces, the force that generally predominantly acts in channels with curvature is the Secondary Dean drag force. Secondary flow arises in the fluid flow through a curved channel because of a mismatch of velocity in the downstream direction between the fluid in the center and near-wall regions of a channel[31,39,72]. Therefore, fluid elements near the channel centerline have larger inertia than fluid near the channel walls and tend to flow outward around a curve, creating a pressure gradient in the radial direction of the channel. Because the channel is enclosed the relatively stagnant fluid near the walls re-circulates inward due to this centrifugal pressure gradient, creating two symmetric vortices[14,37].

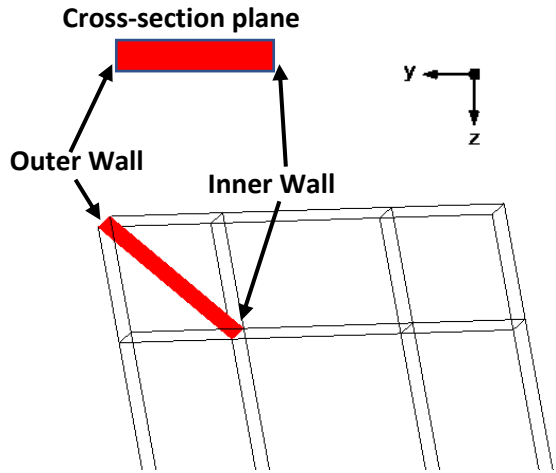

**Figure 17:** Cross-sectional details of the plane along the z-direction in the loop. (z-axis is parallel to the inner and outer wall of the cross-section plane)

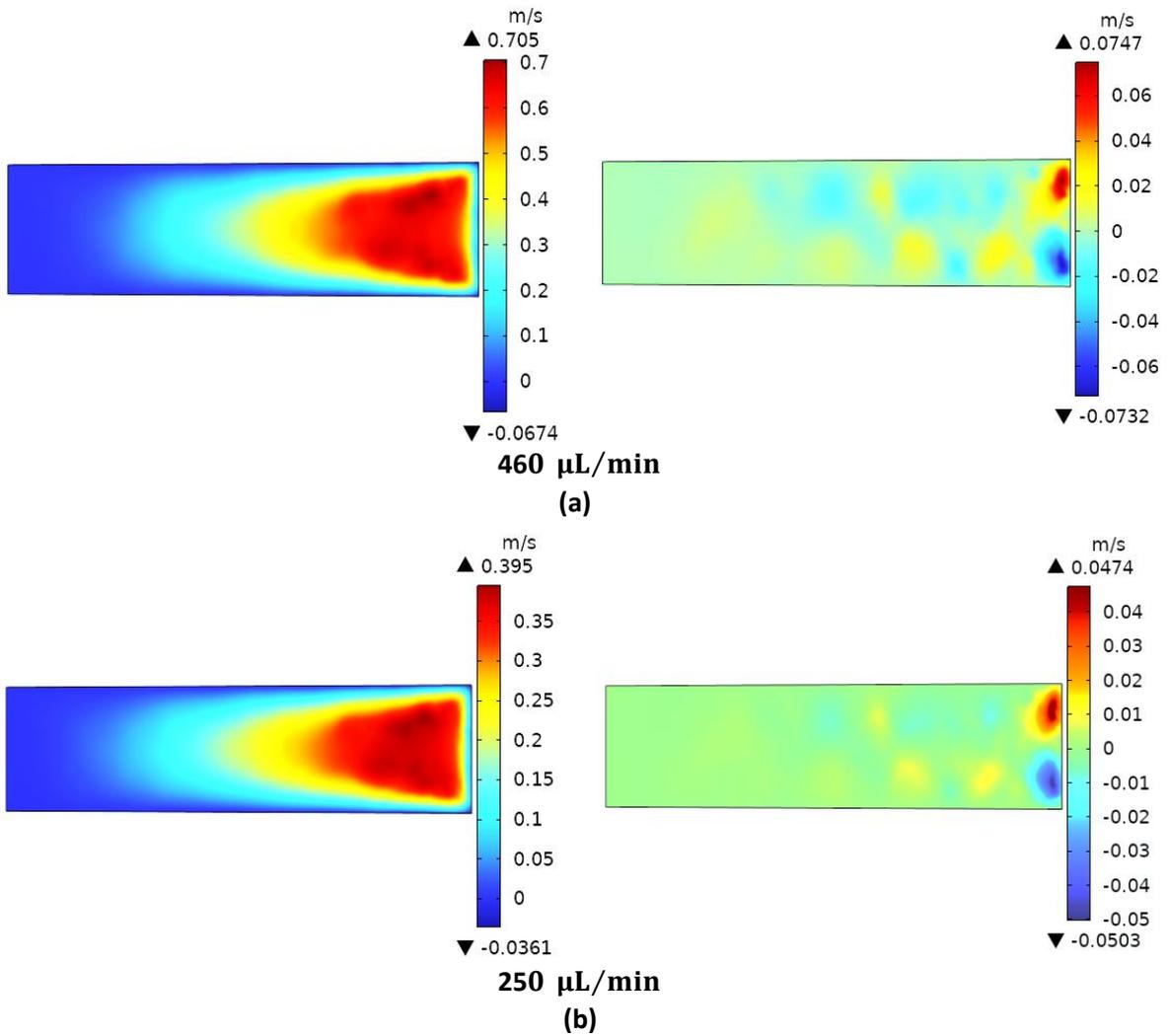

**Figure 18:** x-component (left) and z-component of velocity contour within the microchannel at (a) 460 μL/min (b) 250 μL/min, respectively within the cross-sections depicted in Fig.17.

In serpentine microchannels due to the presence of curvature in the loops, there may be the development of secondary flow, along with the generation of secondary Dean counter-rotating vortices[14,37,43]. This secondary flow accompanied by Dean vortices enhances the mixing effects by agitating the fluid[73,74], which is undesirable for particle focusing in microchannels.

To suppress this mixing effect, a microchannel with a low aspect ratio (AR) is suggested[75]. For channels with small AR the fluid velocity in the z direction is too small to drag particles, hence no outward and inward cross-fluid flow is developed. In this problem formulation, initially, a rectangular microchannel with an AR of 0.25 (channel height (H) 50 $\mu m$ and channel width (w) 200 $\mu m$) is taken. These dimensions are small enough to provide a wide available working area while inhibiting the mixing effects of secondary flow. To demonstrate the effect of secondary flow and the fluid velocity in the z direction, perpendicular cross sections are considered, as shown in Fig. 17, within the 15th loops of the microchannel.

For the present study, three different diameter ratios of particles (n) have been considered viz., 1.2 (7.5 μm & 9 μm), 1.5 (7.5 μm & 11.25 μm), 2 (7.5 μm & 15 μm), where the maximum separation efficiency is found at flow rates of 460 μL/min ($Re_c$ = 152.41), 360 μL/min ($Re_c$ = 119.28) and 250 μL/min ($Re_c$ = 82.83) respectively. As shown in Fig. 18, the secondary flow strengths (z-component) have been compared with the x-component of flow velocity, corresponding to the mentioned flow rates at the chosen cross-section. It has been analyzed, from the simulations, that the magnitude of the z-component of the velocity profile is negligible in comparison to the x-component, which is depicted in Fig. 18. It can be also seen, that no Dean vortices are formed. Consequently, the secondary drag velocity in the z-direction has been neglected. Hence, for the present study, a 2D model geometry with a channel width of 200 $\mu m$ has been used.

## A.4. <u>Validity of one-way coupling and inertial focusing</u>

In the present study, as already mentioned earlier, such particle combinations were analyzed for which, $a/D_h > 0.07$. Hence, the inertial lift forces play a vital role in governing the particle behavior, as suggested in previous works of Di Carlo[14,31,39]. Again for $a/D_h \ll 1$ the underlying fluid flow is not disturbed by the particle motion. Hence, only the particles get affected by the fluid motion. So one-way fluid particle coupling should be able to capture the particle motion within the microchannel. Thus, the condition $0.07 < a/D_h \ll 1$ indicate two phenomena simultaneously.

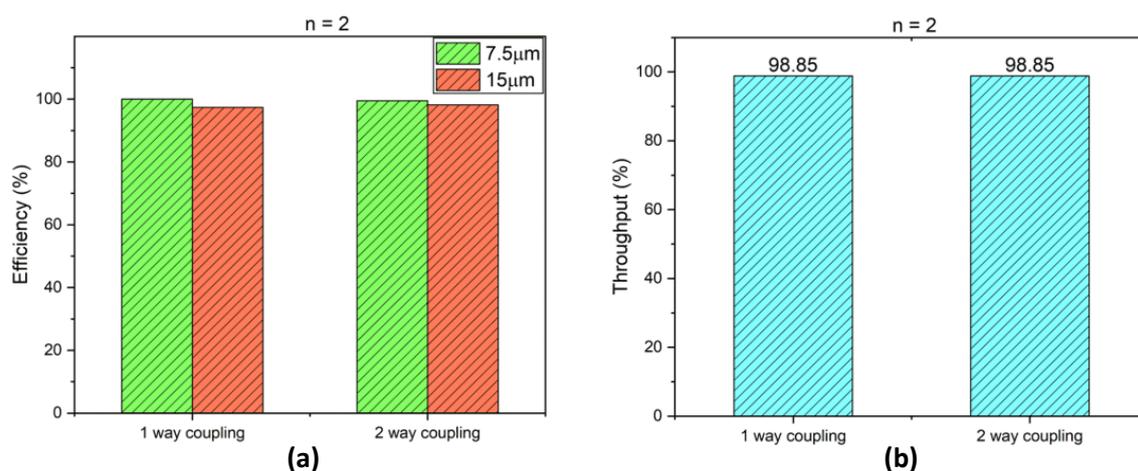

**Figure 19:** (a) Sorting efficiency and (b) throughput comparison for n = 2, at an appropriate flow rate of 250 μL/min, under both one-way and two-way fluid-particle coupling conditions.

The significance and contribution of inertial lift have been discussed in appendix A.2 section. Both one-way and two-way fluid-particle coupling has been performed for n = 2, under an appropriate flow rate of 250 µL/min. The corresponding particle sorting efficiency and throughput at the outlet have been shown in Figs 19(a) and 19(b). As it can be seen from Fig. 19, for both one-way and two-way fluid-particle coupling the particle sorting efficiency and throughput are almost equal. Hence this again, supports the fact that for $a/D_h \ll 1$ as the particle motion does not affect the flow field, both the one-way and two-way coupling produced similar results. Hence, the use of one-way fluid-particle coupling in the present study is justified.

### A.5. Role of other forces:

A.5.1. Effect of virtual mass force:

The virtual mass force also called the added mass force accounts for the additional force required to accelerate a fluid mass surrounding a particle when the particle itself accelerates[76–82]. This effect arises because when a particle moves through a fluid, it must also displace some of the fluid around it, which requires extra force.

Mathematically it is expressed as[76,83–85],

$$F_{\text{vm}} = \frac{1}{12}\pi d_p^3 \rho_l \frac{D}{Dt}(\boldsymbol{u} - \boldsymbol{u_p}) \tag{A1}$$

Now in the present study, to check the effect of virtual mass force, simulation were performed with and without its presence. Corresponding sorting efficiency are plotted which shows no change in result, as shown in Fig. 20(a). Further, representative values of $F_{\text{vm}}$ are computed which is orders of magnitude less than the considered forces viz., Stokes drag, Saffman lift and centrifugal forces, as shown in Fig. 20(b). Thus, the effect of virtual mass force can be omitted in the present work.

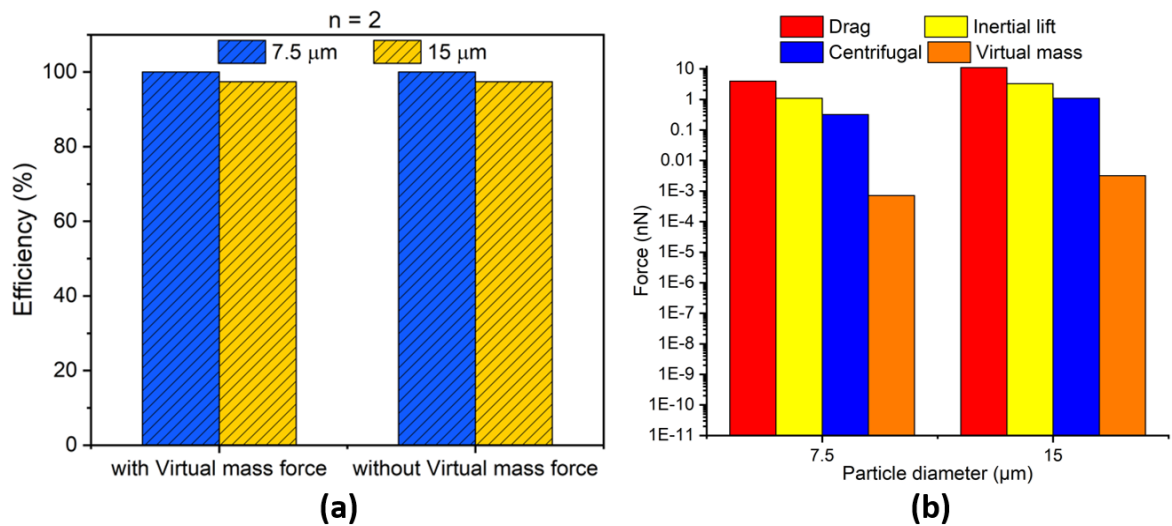

**Figure 20:** (a) Particle sorting efficiency comparison with and without virtual mass force. (b) Magnitude of virtual mass force in logarithmic scale, for n = 2, under effective sorting conditions.

A.5.2. Effect of Archimedes force:

Another force that acts on the particles, during its motion in microfluidics channels, is the Archimedes force, that included both the effect of particle acceleration and gravity[76–79]. Mathematically it is expressed as,

$$F_A = \frac{1}{6}\pi d_p^3 \rho_l \left(\frac{D\boldsymbol{u_p}}{Dt} - \boldsymbol{g}\right) \tag{A2}$$

In the present, the effect of gravity has been neglected, since the ration of gravity force to that of drag force is much smaller than unity.

$$\frac{\rho_p d_p^2 g}{18\mu U} \ll 1 \tag{A3}$$

Thus, similarly the effect of the Archimedes force is also implemented in the model and the corresponding particle sorting efficiency is plotted for n = 2, under its optimized flow conditions, in Fig. 21(a). There is no efficiency change in the two conditions. Furthermore, Fig 21(b) depicts, representative values for $F_A$, which is orders of magnitude less than the other forces considered for the present work. Hence, we can safely neglect the effect of Archimedes force for the present study.

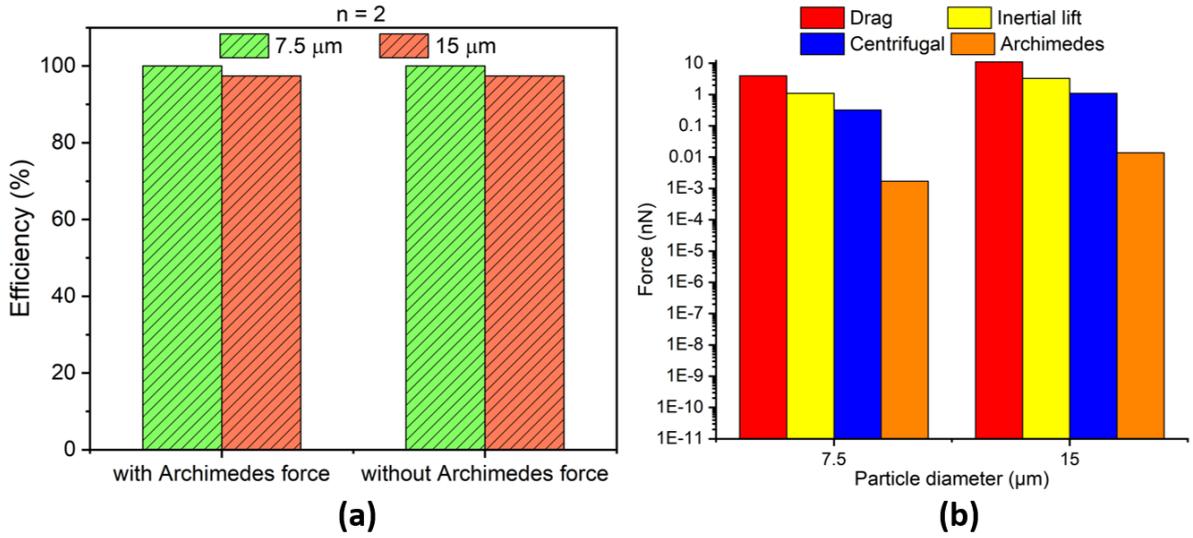

**Figure 21:** (a) Particle sorting efficiency comparison with and without Archimedes force. (b) Magnitude of Archimedes force in logarithmic scale, for n = 2, under effective sorting conditions.

A.5.2. Effect of Basset force:

The Basset force describes the force caused by the temporal delay in boundary layer development and viscous effects when an acceleration motion of particles in fluid is in existence[85–87]. As the particle acceleration affects its surrounding fluid and causes the added mass force, the viscous effects of the surrounding fluid also effects the accelerating particle under the Basset force force[85,88–90].

Mathematically it can be expressed as,

$$F_{Ba} = \frac{3}{2} d_p^2 \sqrt{\pi \rho_f \mu_f} \int_{-\infty}^{t} \frac{\frac{d}{d\tau}(u_f - u_p)}{\sqrt{t-\tau}} d\tau. \tag{A4}$$

Most of the works, have neglected the effect of Basset force, due to the dominance of the drag force over it, by orders of magnitude, and absence of flow perturbation[83,84,87,91–94]. Furthermore, in the present work, since the phase velocity relaxation length scale (~$10^{-10}$) is much smaller than the length scale of the problem (~$10^{-6}$), the Basset force can be neglected[77,87,91,95].

The relative importance of the Basset force, compared to the drag is governed by the parameter $\varepsilon^2$, defined as[87,96],

$$\varepsilon^2 = \frac{9m_a}{m_a+m_p} \quad (A5)$$

where $m_p$ is the particle mass and $m_a$ is the virtual (added) mass which accounts for the inertia added to the system, due to the accelerating body inside the fluid. It can be shown[96] that for the spherical particles, the added mass can be calculated by $m_a = \frac{2}{3}\rho_f \pi r_p^3$. By substitution, the $\varepsilon^2 \to 0$, which demonstrates the dominancy of the drag force over the Basset force, and, thus we can neglect the Basset force.

### A.6. Particle-wall interaction:

Regarding the particle-wall interaction modelling methodology implemented in this paper, the specular reflection model is utilized. This model not only prevents the occurrence of a zero-gap condition between the particle and the wall but also conserves the particle's kinetic energy, ensuring a physically accurate and robust implementation of the bounce boundary condition. Rather than allowing the particle to reach a regime where the drag force becomes singular due to vanishing separation from the wall, the model detects the impact through trajectory interpolation. At the moment of impact, the drag wall correction model is not applied. Instead, the specular reflection model treats the interaction as a perfectly elastic collision, thereby avoiding the singularity in the drag force[54,62]. Furthermore, in this model, the particle's trajectory is tracked using discrete time-stepping. The position $x_p(t)$ at a time instant is updated based on its velocity and external forces. At each time step, the algorithm checks if the predicted position $x_p(t + \Delta t)$ crosses the wall. If a particle is detected inside the wall, interpolation is used to estimate the exact collision time[97–99]. The specular reflection model is commonly used for elastic collisions, where the normal velocity component is reversed while the tangential component is unchanged[100,101]

$$\boldsymbol{v_p}^* = \boldsymbol{v_p} - 2(\boldsymbol{v_p} \cdot \mathbf{n})\mathbf{n} \quad (A6)$$

Since the current Finite Element solving approach tracks the particle motion using finite time steps[54], it does not check for zero gap at every instant. If a particle trajectory is detected to intersect the wall between two time-steps, the software solves for the precise collision time using interpolation[54]. Collision is detected before the singularity occurs. Instead of letting the drag force grow uncontrollably, the particle is reflected back based on kinematic principles, as shown in Fig. 22. Even though wall corrections lead to infinite drag as the particle approaches the wall, the current modelling approach prevents this from happening by enforcing reflection before that point. The momentum-based bounce condition does not rely on the drag model, so the particle does not experience an artificially large drag force during impact. This allows realistic particle trajectories while avoiding numerical instability.

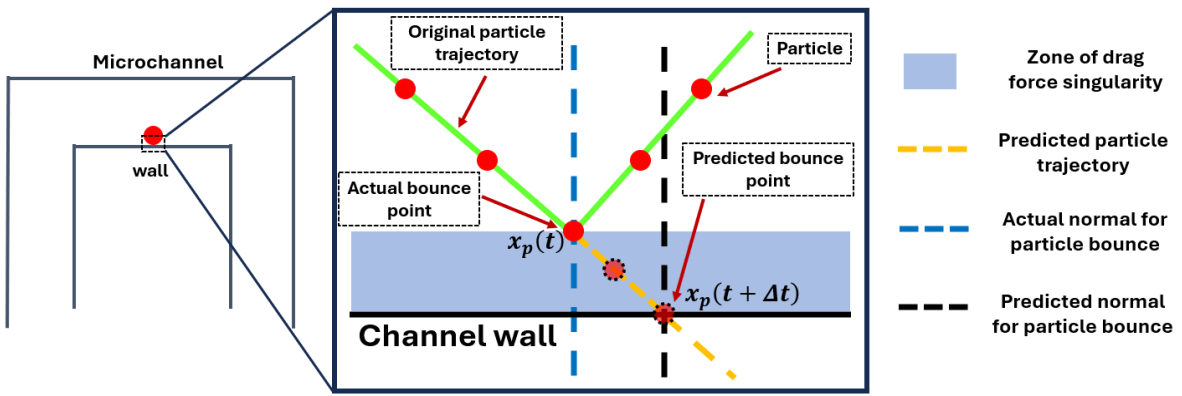

**Figure 22:** Illustration for the particle specular reflection model.